\documentclass[11pt,a4paper]{article}

\usepackage{cite}
\usepackage{amsmath, amssymb}
\usepackage{graphics}
\graphicspath{{figures/}}
\usepackage{subfigure}
\usepackage{graphicx}
\usepackage{amsfonts}
\usepackage{longtable}
\usepackage{bm}
\usepackage{epsf, epsfig}
\usepackage{epstopdf}
\usepackage{algorithm}
\usepackage{algorithmic}
\usepackage{color}
\usepackage{array}
\usepackage{float}
\usepackage{enumerate}
\usepackage{fontenc}
\usepackage{parskip}
\usepackage{fixltx2e}
\usepackage{multirow}
\usepackage{times}
\usepackage{amsmath}
\usepackage{amssymb}
\usepackage{booktabs}

\usepackage[left=2.5cm,top=3cm,right=2.5cm,bottom=3cm,bindingoffset=0.5cm]{geometry}
\UseRawInputEncoding
\begin{document}
\onecolumn

\title{\Large Structured Sparsity Modeling for Improved Multivariate Statistical Analysis based Fault Isolation}

\author{{\normalsize Wei Chen$^{1}$, \footnote{Corresponding author:
Jiusun Zeng: jszeng@cjlu.edu.cn} Jiusun Zeng$^{1}$, Xiaobin Xu$^{2}$, Shihua Luo$^{3}$, Chuanhou Gao$^{4}$}
\\
{\scriptsize $^1$ College of Metrology and Measurement Engineering, China Jiliang University, Hangzhou 310018, P.R. China.}
\\
{\scriptsize $^2$ School of Automation, Hangzhou Dianzi University, Hangzhou 310018, P. R. China.}
\\
{\scriptsize $^3$ School of Statistics, Jiangxi University of Finance and Economics, Nanchang 330013, P. R. China.}
\\
{\scriptsize $^4$ School of Mathematical Sciences, Zhejiang University, Hangzhou 310027, P. R. China.}
\date{\empty}
}
\maketitle

\date{}

\maketitle

\begin{abstract}
In order to improve the fault diagnosis capability of multivariate statistical methods, this article introduces a fault isolation framework based on structured sparsity modeling. The developed method relies on the reconstruction based contribution analysis and the process structure information can be incorporated into the reconstruction objective function in the form of structured sparsity regularization terms. The structured sparsity terms allow selection of fault variables over structures like blocks or networks of process variables, hence more accurate fault isolation can be achieved. Four structured sparsity terms corresponding to different kinds of process information are considered, namely, partially known sparse support, block sparsity, clustered sparsity and tree-structured sparsity. The optimization problems involving the structured sparsity terms can be solved using the Alternating Direction Method of Multipliers (ADMM) algorithm, which is fast and efficient. Through a simulation example and an application study to a coal-fired power plant, it is verified that the proposed method can better isolate faulty variables by incorporating process structure information.

{\small \textbf{Keywords:} Fault isolation; Structured sparsity; Multivariate statistical analysis; ADMM.}
\end{abstract}

\section{Introduction}
With the rapid advancement of sensing and instrumentation technology, the amount of data collected and analyzed in modern industrial plants has grown exponentially. The easy availability of big datasets has made data-driven process monitoring and fault diagnosis an important tool in ensuring safety and reliability of industrial processes. The last few decades have witnessed a mushrooming of research works in this field, which have been summarized in a series of review articles~\cite{Macgregor2012,Yin2014,Ge2017}. The abundance of data information greatly facilitates the task of process monitoring, however, it also poses challenges due to issues like high dimensionality, multi-scale, inconsistent data quality etc. To handle big datasets in large-scale processes, different kinds of distributed process monitoring methods have been researched, such as distributed principal component analysis(PCA)~\cite{Ge2013} and distributed canonical correlation analysis(CCA)~\cite{Chen2019}. Distributed process monitoring methods performs fault detection by decomposing the process into a sets of subprocesses~\cite{Jiang2019}, and perform fault isolation and localization using traditional methods like contribution plots. They have proved to be effective in fault detection. In fault isolation and localization, however, distributed methods suffer from the same weaknesses as traditional methods, e.g., the ``smearing effect''~\cite{Fuentes2018} and dilution effect for localized and incipient fault~\cite{Reis2019}. Whilst the ``smearing effect'' can be alleviated via appropriate variable selection~\cite{Kuang2015,Zhao2016} or risk minimization~\cite{Zheng2016}, it is shown in Ref.~\cite{Kerkhof2013} that contribution plots are efficient only when physical insight is available. This is especially true for large-scale processes, as more variables and data samples will aggravate the dilution effect and compromise the accuracy of fault isolation.

In order to improve the performance of fault isolation and diagnosis, process structure based approaches have received significant attention~\cite{Reis2019}. By integrating structure information like correlation and causality into the model, process structure based approaches are able to produce fault isolation results that are more in line with practical production conditions. Such structure information can be learned from data using Granger causality analysis~\cite{Yuan2014}, transfer entropy~\cite{Duan2013} or obtained from background and expert knowledge~\cite{Xiao2018}. Based on the structure information, a straightforward idea is to incorporate it into traditional multivariate statistical methods using, for example, plug-in approach~\cite{Rato2017}, hierarchical approach~\cite{Luo2017} or two-step methods which apply multivariate statistical methods for fault detection and causality analysis for fault isolation and diagnosis~\cite{Gharahbagheri2017}. Another way to utilize the process structure is through the graph network models like Bayesian network~\cite{Cai2017} or structured graph models~\cite{Xia2015,Zeng2017}. In addition, other structured network approaches~\cite{Rato2015,Charbonnier2010} can also use the process structure to diagnose fault by monitoring the changes in the networked structure or the strength of mutual links. Although the literature reported successes in many applications, challenges still remain. For example, structure learning methods based on Granger causality analysis and transfer entropy are still data-driven and hence difficult to incorporate qualitative process knowledge. Graph network models work well in small-scale systems as the network structure can be easily analyzed. For large-scale systems, however, a complete graph network structure can be difficult to obtain, there still lacks approaches to effectively utilize incomplete network structure.

More recently, it is found that sparsity is commonly observed in the fault structure of industrial processes, as most process faults are localized and affect only a subset of process variables. This finding has inspired the application of sparse methods in fault isolation, including those based on shrinking PCA~\cite{Xie2013}, LASSO(Least absolute shrinkage and selection operator)~\cite{Yan2015} and sparse canonical variate analysis~\cite{Lu2018}. By introducing sparsity regularization terms such as $l_1$(sum of absolute value of all elements of a vector or matrix) and $l_{2,1}$(sum of the Euclidean norms of the rows of a matrix) in the fault isolation objective function, the hidden fault structure can be better revealed so that the sparse methods are less apt to the ``smearing effect''. However, for highly correlated process variables, it is revealed in~\cite{Zhao2006} that the LASSO-like problem will violate the \emph{irrepresentable condition}, which indicates that it cannot recover the true sparsity structure when there is colinearity between process variables that are assumed to be irrelevant due to the existence of noises and disturbances, resulting in decreased isolation accuracy. In order to improve the isolation accuracy, a natural extension is to integrate sparsity with process structure so that \emph{structured sparsity} can be achieved, for example, in the form of graph Laplacian matrix~\cite{Liu2019} or group LASSO~\cite{Shang2019}. The improved fault isolation performance in Refs.~\cite{Liu2019} and~\cite{Shang2019} highlighted the advantages of introducing structured sparsity in fault isolation: i) The candidate faulty subspace can be reduced by incorporating a priori information on process/fault structure using structured sparsity terms; ii) The interpretability of fault isolation results can be significantly improved by incorporating process structure into the sparsity regularization terms; iii) The irrepresentable condition can be better met and incomplete structure information can be utilized.

Despite the advantages brought by structured sparsity, the research on this subject is still limited. In this paper, a structured sparsity modeling framework for multivariate statistical analysis methods is proposed by considering a range of structure information in the form of partially known sparse support, block sparsity, clustered sparsity and tree-structured sparsity. Such structure information can be included in the fault isolation objective function using different structured sparsity regularization terms. In addition, an effective optimizing routine based on alternating direction method of multipliers(ADMM)\cite{Boyd2010} is proposed, which can be easily extended to a parallel or distributed version to accommodate big datasets. The goal of this paper is to provide a practical guide on how to utilize different kinds of structure information in process industry to enhance the fault isolation performance.

\section{Problem formulation}
\subsection{Monitoring statistics for fault detection}
Let $\mathbf{X}\in \mathbb{R}^{n \times m}$ be a data matrix storing $n$ samples of $m$ process variables, denote a sample to be $\mathbf{x}\in \mathbb{R}^m$ and $x_i (i=1,\dots,m)$ to be the $i$th variable. Assume that $\mathbf{x}$ is normalized to have a zero mean and unit variance, according to Ref.~\cite{Yan2015}, typical multivariate statistical methods perform decomposition of the data sample as follows.
\begin{eqnarray}\label{eqn1}
\mathbf{x}= \mathbf{As}+\mathbf{e}
\end{eqnarray}
Here $\mathbf{s}=\mathbf{w}^T\mathbf{x}$ is the projection of the sample $\mathbf{x}$ in the $l(l \leq m)$ dimensional subspace and $\mathbf{e}=(\mathbf{I}-\mathbf{Aw}^T)\mathbf{x}$ is the residual, with the projection matrix $\mathbf{w}\in \mathbb{R}^{m\times l}$, loading matrix $\mathbf{A}\in \mathbb{R}^{m\times l}$ and the identity matrix $\mathbf{I}$. Such decomposition can be achieved by maximizing variance(principal component analysis), negentropy(independent component analysis, ICA) or correlation(canonical correlation analysis, CCA). Based on the decomposition in Eq.(\ref{eqn1}), monitoring statistic can be defined in the lower dimensional space to detect process fault as follows.
\begin{eqnarray}\label{eqn2}
\text{T}^2= \mathbf{s}^T \mathbf{s}=\mathbf{x}^T \mathbf{ww}^T \mathbf{x}
\end{eqnarray}
And the SPE statistic is defined in the residual space as
\begin{eqnarray}\label{eqn3}
\text{SPE}=\mathbf{e}^T\mathbf{e}=\mathbf{x}^T (\mathbf{I}-\mathbf{Aw}^T)^T(\mathbf{I}-\mathbf{Aw}^T)\mathbf{x}
\end{eqnarray}
Both the statistics can be written in a compact form as
\begin{eqnarray}\label{eqn4}
R=\mathbf{x}^T \mathbf{Mx}
\end{eqnarray}
Here $R$ corresponds to quadratic statistics like Hotelling's $T^2$ or SPE, $\mathbf{M}$ is a symmetric and positive semi-definite matrix. When used for fault detection, corresponding control limits can be obtained based on samples collected under normal operation conditions using statistical theory or empirical analysis. For a specific test sample, violation of either statistic indicates that it is faulty.
\subsection{Sparse reconstruction for fault isolation}
Once a fault arises, it is important to accurately identify and isolate the faulty variables. In order to quantify the contribution of each variable to the fault, a natural idea is to reconstruct the ``true value'' by removing the effect of fault. For a faulty sample $\mathbf{x}$, it can be decomposed into the sum of the ``true value'' $\mathbf{x}^*$ and a fault vector $\mathbf{f}$ as follows.
\begin{eqnarray}\label{eqn5}
\mathbf{x}=\mathbf{x}^*+\mathbf{f}
\end{eqnarray}
The fault vector can be reconstructed by solving the following optimization problem
\begin{eqnarray}\label{eqn6}
\tilde{\mathbf{f}}=\mathop {\arg \min }\limits_{\mathbf{f}} (\mathbf{x}-\mathbf{f})^T \mathbf{M}(\mathbf{x}-\mathbf{f})
\end{eqnarray}
Here $\tilde{\mathbf{f}}$ is the estimation of the true fault vector. In most cases, the fault only affects a subset of variables, this is especially true for large-scale processes. Following Ref.~\cite{Yan2015}, Eq.(\ref{eqn6}) can be modified to achieve sparsity by imposing the $l_1$ regularization term on the fault vector.
\begin{eqnarray}\label{eqn7}
\tilde{\mathbf{f}}=\mathop {\arg \min }\limits_{\mathbf{f}} (\mathbf{x}-\mathbf{f})^T \mathbf{M}(\mathbf{x}-\mathbf{f})+\lambda \|\mathbf{f}\|_1
\end{eqnarray}
Here $\lambda$ is the weight on the $l_1$ term. Eq.(\ref{eqn7}) can be solved using methods like block coordinate reduction (BCD)~\cite{Sum2020}. By appropriately selecting the weight, a sparse fault vector can be estimated. The nonzero elements in $\tilde{\mathbf{f}}$ correspond to faulty variables, whilst zero elements relate to normal variables. Achieving sparsity in the reconstructed fault vector can effectively reduce the number of variables to be attended by operators in correcting the fault. Hence it is advantageous to traditional contribution plots, which tend to produce a fault vector without zero element. However, like traditional contribution plots, the sparse reconstruction problem in Eq.(\ref{eqn7}) does not consider process structure information. More importantly, if the process variables are highly correlated, the design matrix of LASSO-like problem will violate the \emph{irrepresentable condition}~\cite{Zhao2006} and this will negatively affect the fault isolation accuracy.

\section{Physics based structured sparsity for fault isolation}
\begin{figure}[h]
\begin{center}
\includegraphics[width=12cm]{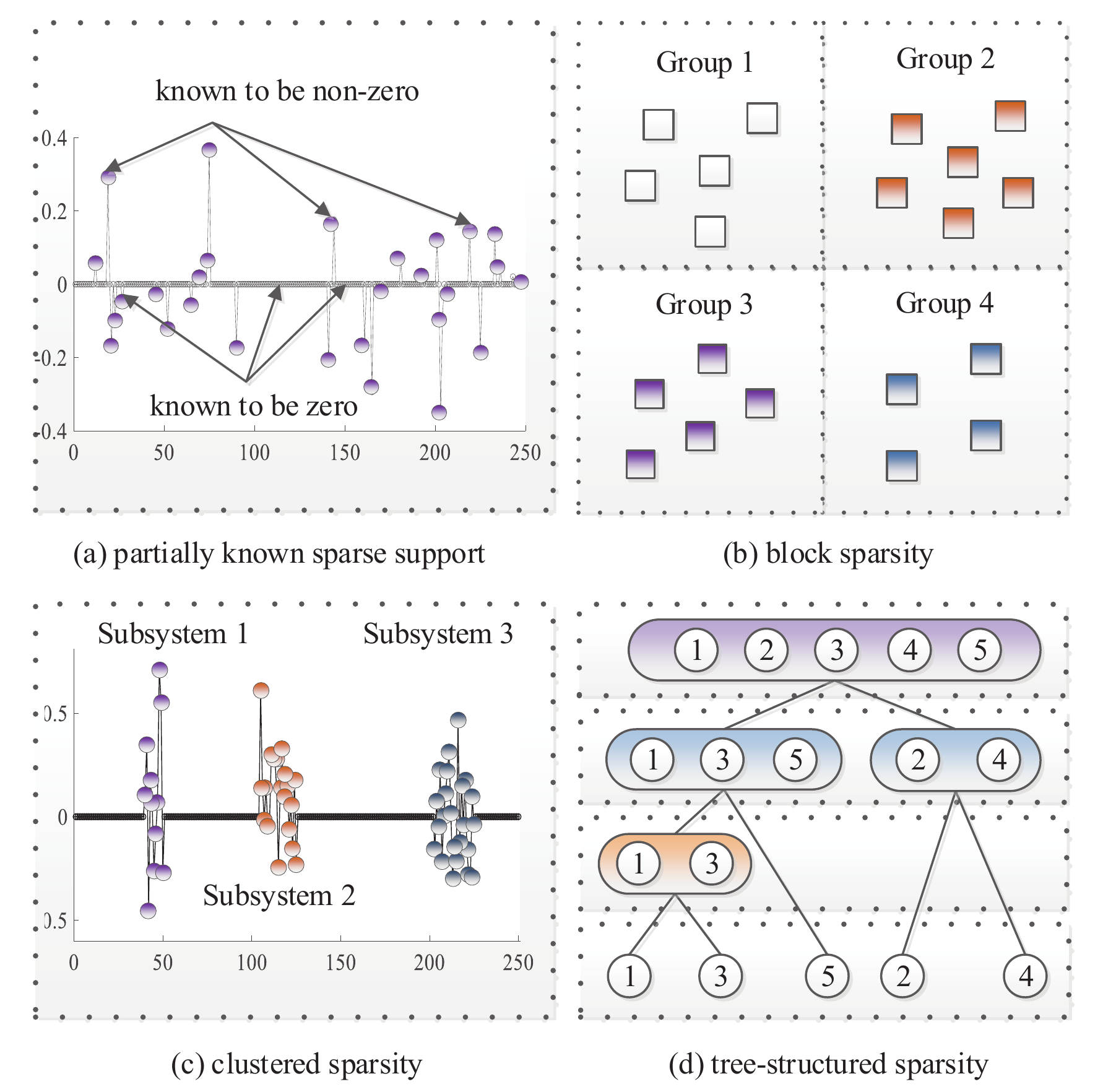}
\setlength{\abovecaptionskip}{-3pt}
\caption{Several common process/fault structures and their structured sparse representation: (a)Prior knowledge about some variables being normal/abnormal; (b)Naturally occurring variable modules or blocks; (c)Variables having similar fault behavior in a clustered way; (d)Tree-structured correlation/causal relationship between variables.}  \label{Figure1}
\end{center}
\end{figure}

Physical structure in industrial processes can be used to improve the performance of monitoring and control~\cite{Wu2020}. This section introduces four structured sparse characteristics of process/fault structures, i.e., partially known sparse support, block sparsity, clustered sparsity and tree-structured sparsity, as shown in Figure~\ref{Figure1}.
These regularization alternatives are suitable for different scenarios. For example, the partial known sparse/non-sparse support can be used when there is limited information on the fault structure and it is known that some of the process variables are or are not related with the fault. Block sparsity regularization can be applied when there is known information on the block structure of process variables. If more information is available and the tree-structure of the system can be obtained, the tree-structured sparsity can be used. Hence the application of these regularization alternatives depends on which kind of structured information is available. The partial known sparse/non-sparse support can be applied together with block and tree-structured sparsity if both kinds of structure information are available. For block and tree-structured sparsity, they are generally not used simultaneously as tree structure can be seen as an advanced kind of block structure.
\subsection{Partially known sparse support}
The problem of partially sparse recovery has been studied in the field of comprehensive sensing~\cite{Vaswani2010}, in which prior knowledge on the support of a signal may exist. The concept of support is widely used in mathematics. In the background of fault isolation, a sparse support is a subset of normal variables whose contribution to the fault is zero. Conversely, a non-sparse support is a subset of faulty variables whose contribution to the fault is non-zero. In process monitoring applications, such prior knowledge is commonly available. Under a faulty scenario, the operators may have high confidence that certain variables are not affected by the fault. These variables should have no contribution to the fault and consist of the sparse support in fault isolation. Introducing sparse support can reduce the variable space to be searched, and is helpful for improving the fault isolation accuracy and reducing computation burden for isolation of localized fault in large-scale systems. In this subsection, only the partially known sparse support is considered and the partially known non-sparse support can be dealt with in a similar way.

Let $\mathbf{T}\subset \left\{1,2,\dots,d \right\}, d<m$ be the known sparse support of the fault vector and the following holds
\begin{eqnarray}
\mathbf{f}_g = 0 \qquad \forall g \in \mathbf{T}
\end{eqnarray}
Ref.~\cite{Vaswani2010} proved that if partial support of the signal is known, a more accurate reconstruction at the same measurement rate can be obtained by constraining the $l_1$ norm of the signal on the complement of the known support. Considering the sparse support, the fault isolation problem in Eq.(\ref{eqn7}) can be modified to the following optimization problem.
\begin{eqnarray}\label{eqn9}
\tilde{\mathbf{f}}=\mathop {\arg \min }\limits_{\mathbf{f}} (\mathbf{x}-\mathbf{f})^T \mathbf{M}(\mathbf{x}-\mathbf{f})+\lambda \|(\mathbf{f})_{\mathbf{T}_c}\|_1
\end{eqnarray}
Here $\mathbf{T}_c$ is the complement of $\mathbf{T}$. The above optimization problem now becomes to find the faulty variables outside of $\mathbf{T}$ that satisfy the data constraint. In this way, the faulty space to be searched will be limited and the computation load will be reduced, resulting in more accurate fault isolation.

\subsection{Block sparsity}
Another way of incorporating process structure is to consider the natural blocks or modules of variables. In industrial production practice, a large-scale system can be decomposed into multiple local sub-blocks, which are connected by multiple streams. Therefore, variables can be divided into several blocks, and highly correlated variables will naturally fall into the same group and show similar contributions. If the block structure is incorporated into the objective function as a priori information, naturally, the accuracy, practicality and efficiency of fault diagnosis can be improved. This has inspired the application of methods like group LASSO~\cite{Shang2019} in fault isolation.

Assume the $m$ process variables fall into $b$ non-overlapping blocks and $\mathbf{x}=[\mathbf{x}_1 \ \mathbf{x}_2 \dots \mathbf{x}_b]^T$. Let $p_l$ denote the number of variables in the $l$th block, so that $\sum \limits_{l=1}^b p_l=m$. Following Shang \emph{et. al}~\cite{Shang2019}, the group LASSO based fault isolation problem can be formulated as follows.
\begin{eqnarray}\label{eqn10}
\tilde{\mathbf{f}}=\mathop {\min }\limits_{\bf{f}} {\left( {{\bf{x}} - {\bf{f}}} \right)^T}{\bf{M}}\left( {{\bf{x}} - {\bf{f}}} \right) + \lambda \sum\limits_{l = 1}^b {\sqrt {{p_l}} {{\left\| {{{\bf{f}}_l}} \right\|}_2}}
\end{eqnarray}
where ${{\bf{f}}_{l}}$ corresponds to the $l$th block of variables in ${\bf{f}}$ and $\sum\limits_{l = 1}^b {\sqrt {{p_l}} {{\left\| {{{\bf{f}}_{l}}} \right\|}_2}}$ is called the block $l_{2,1}$ penalty term, which can be regarded as an intermediate between the ${l_1}$ norm penalty and ${l_2}$ norm penalty. With appropriate $\lambda$, the solution of Eq.(\ref{eqn10}) will produce ${{{\bf{f}}_l}}$ that is either zero or nonzero. Therefore, faulty blocks can be identified and the root cause of the fault needs to be located according to the contribution of each variable in the block. It should be noted that overlapping blocks can also be allowed~\cite{Yuan2011}, for the sake of simplicity, however, it is not considered here.

Group LASSO is effective for small-scale problem, for large-scale system, however, the block size may be large and the localization of root cause may still be difficult. Hence it is desired to achieve sparsity also in the faulty blocks to facilitate fault localization. This can be achieved by modification of Eq.(\ref{eqn10}) to the so-called sparse group LASSO~\cite{Simon2013} as follows.
\begin{eqnarray}\label{eqn11}
\widetilde {\bf{f}} = \mathop {\min }\limits_{\bf{f}} {\left( {{\bf{x}} - {\bf{f}}} \right)^T}{\bf{M}}\left( {{\bf{x}} - {\bf{f}}} \right) + \left( {{\rm{1 - }}\alpha } \right)\lambda \sum\limits_{l = 1}^b {\sqrt {{p_l}} {{\left\| {{{\bf{f}}_l}} \right\|}_2}} {\rm{ + }}\alpha \lambda {\left\| {\bf{f}} \right\|_{\rm{1}}}
\end{eqnarray}
where $\alpha  \in \left[ {{\rm{0,1}}} \right]$ is the weighting parameter. In Eq.(\ref{eqn11}), the first penalty term is to ensure block-wise sparsity and the second term is used to control sparsity within groups. Introducing sparsity inside the faulty blocks will help reduce the faulty variables to be attended and hence is helpful for localization of fault root cause. Similarly, Eq.(\ref{eqn11}) can also be extended to handle overlapping blocks by applying a temporary weighting scheme on the regularization weights.
\subsection{Clustered sparsity}
In some cases, limited information on variable blocks or modules is available, there is a need for clustering the variables outside the natural blocks into new blocks. This is a necessary step for highly correlated variables outside the natural blocks to meet the \emph{irrepresentable condition}~\cite{Zhao2006}. Different kinds of clustering methods can be applied, for example those based on correlation coefficient~\cite{Ge2013}, mutual information~\cite{Jiang2014} and partial correlation analysis~\cite{De2004,Reis2013}. For simplicity, it is assumed all the variables have been clustered into a total of $q$ groups as $\mathbf{S} = \{ {S_1},{S_2}, \ldots ,{S_q}\} $, with $S$ being the set of all variables.

Once a fault is detected, clustered sparsity can be incorporated into the fault isolation objective as follows.
\begin{eqnarray}\label{eqn12}
\tilde{\mathbf{f}}=\mathop {\min }\limits_{\bf{f}} {\left( {{\bf{x}} - {\bf{f}}} \right)^T}{\bf{M}}\left( {{\bf{x}} - {\bf{f}}} \right) + \lambda \sum_{j=1}^q\|\mathbf{f}_{{S}_j}\|_2
\end{eqnarray}
Here $\mathbf{f}_{{S}_j}$ corresponds to the fault vector of all variables falling into ${{S}_j}$. Eq.(\ref{eqn12}) also allows partial clustering, with a part of variables not belonging to any group, so that Eq.(\ref{eqn12}) can be modified as
\begin{eqnarray}\label{eqn13}
\tilde{\mathbf{f}}=\mathop {\min }\limits_{\bf{f}} {\left( {{\bf{x}} - {\bf{f}}} \right)^T}{\bf{M}}\left( {{\bf{x}} - {\bf{f}}} \right) + \lambda_1 \sum_{j=1}^q\|\mathbf{f}_{{S}_j}\|_2+\lambda_2{\left\| {{{\bf{f}}_{{\mathbf{S}_c}}}} \right\|_1}
\end{eqnarray}
where $\mathbf{S}_c$ is the complement of $\mathbf{S}$. Solution of Eqs.(\ref{eqn12}) and (\ref{eqn13}) can be obtained in a similar way to the case of block sparsity as they share similar model structure.

\subsection{Tree-structured sparsity}
Process variables in industrial systems often exhibit networked or tree-structured correlation/causal characteristics. Such networked or tree structure can be useful for fault isolation and propagation analysis. A tree can be generated through process knowledge. When process knowledge is not available, it can also be learned from data using methods like hierarchical clustering~\cite{Nazari2015} or score-based approaches~\cite{Chickering2000}. This subsection considers the incorporation of tree-structured sparsity in the fault isolation problem, in the hope of obtaining improved fault isolation accuracy and clearer fault propagation path.

\begin{figure}[htbp]
\begin{center}
\includegraphics[width=12cm]{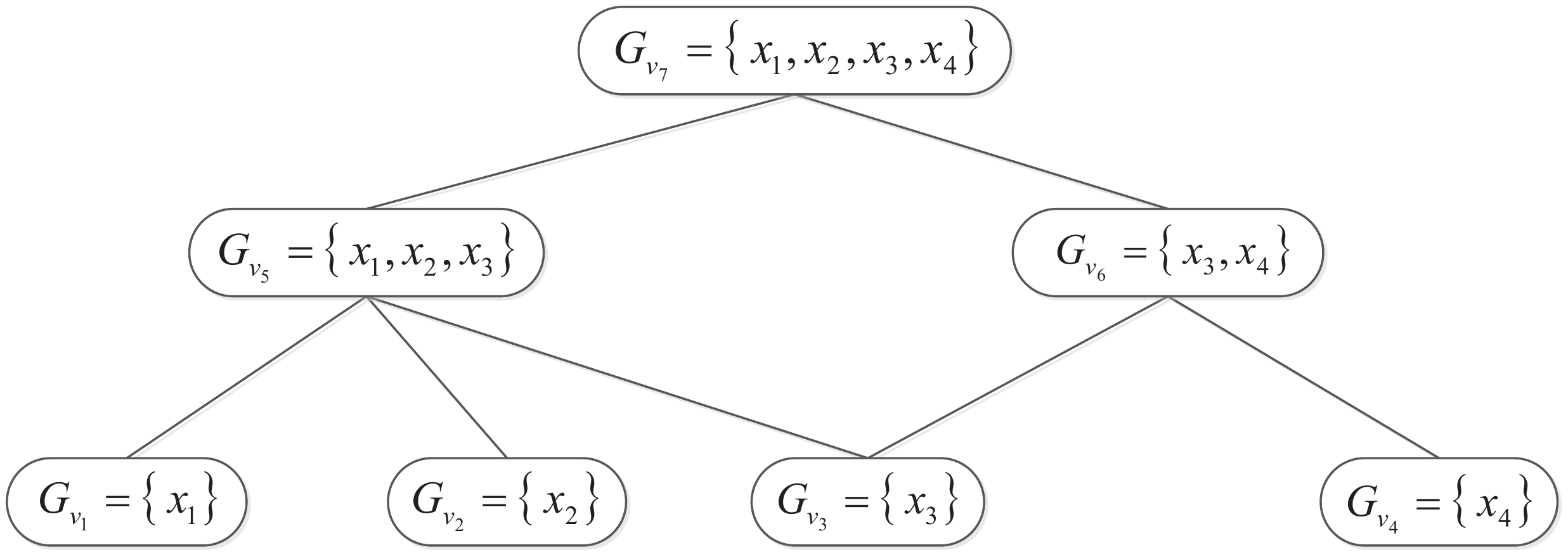}
\caption{A typical four-variable system with tree structure.}  \label{Figure2}
\end{center}
\end{figure}
Assume the relationship between process variables can be represented as a tree $T$ whose set of vertices $V$ has a size of $M$, with $v\in V$ being its node. Figure \ref{Figure2} shows a tree-structured system with four variables, note that the variables can be overlapped in the internal nodes. In Figure \ref{Figure2}, each of the four leaf nodes corresponds to a process variable and each internal nodes associates to a grouping of process variables. The closer the internal nodes is to the leaf nodes the higher correlation exists in the group. Let $G_v$ denote the group of variables in node $v$, the tree can be expressed as $T=\{G_{v_1},G_{v_2},\dots,G_{v_M}\}$, with $M$ being the number of groups. It should be noted that each leaf node is considered to be a group so that the number of groups is equivalent to the number of nodes. It can be seen that there are a total of 7 nodes/groups in Figure \ref{Figure2}.

Based on the tree structure, the fault isolation problem in Eq.(\ref{eqn7}) can be modified as follows~\cite{Kim2010}.
\begin{eqnarray}\label{eqn14}
\tilde{\mathbf{f}}=\mathop {\arg \min }\limits_{\mathbf{f}} (\mathbf{x}-\mathbf{f})^T \mathbf{M}(\mathbf{x}-\mathbf{f})+\lambda \sum_{j=1}^M \omega_{v_j}\|\mathbf{f}_{G_{v_j}}\|_2
\end{eqnarray}
Here $\mathbf{f}_{G_{v_j}}$ is the set of coefficients assigned to $\mathbf{f}$ by the variables in node/group $G_{v_j}$, weight $\omega_{v_j}$ reflects the strength of correlation within group $v_j$.

Let $W(v_{\text{root}})=\lambda \sum_{j=1}^M \omega_{v_j}\|\mathbf{f}_{G_{v_j}}\|_2$, $W(v_{\text{root}})$ represents the penalty accumulated at the root node after recursion according to the tree structure. Eq.(\ref{eqn14}) can now be expressed as follows.
\begin{eqnarray}\label{eqn15}
\tilde{\mathbf{f}}=\mathop {\arg \min }\limits_{\mathbf{f}} (\mathbf{x}-\mathbf{f})^T \mathbf{M}(\mathbf{x}-\mathbf{f})+W(v_{\text{root}})
\end{eqnarray}
It should be noted that $W(\cdot)$ can be defined on any node.
\begin{equation}
  \label{eqn16}
W\left( v_j \right) = \left\{ \begin{array}{l}
\sum\limits_{{M_c} \in {G_{v_j}}} {\| {{{\bf{f}}_{{M_c}}}} \|_1} {\rm{ }} \qquad \qquad\quad \qquad \ \ \ {\text{if $v_j$ is a leaf node}}\\
{s_{v_j}} \cdot \sum {\left| {W\left( c \right)} \right| + {g_{v_j}} \cdot {{\left\| {{{\bf{f}}_{{G_{v_j}}}}} \right\|}_2}{\rm{ }}} \qquad {\text{otherwise}}
\end{array} \right.
\end{equation}
Here $W(c)$ is the child node of $W(v_j)$. The weights $s_{v_j}$ and $g_{v_j}$ are introduced for each internal node and the root node, with $s_{v_j}+g_{v_j}=1$. The weight $g_{v_j}$ on the $l_2$ norm determines how much weight is put on jointly selecting the variables in the associated node, whilst the weight $s_{v_j}$ determines how much weight is put on selecting each child node separately.

Take the tree structure in Figure~\ref{Figure2} for example, the penalty terms $W(\cdot)$ can be defined by the following steps.
\begin{equation}
  \label{eqn17}
\begin{array}{l}
W\left( {{v_1}} \right) = \| {{{\bf{f}}_1}} \|_1 \qquad\qquad\qquad W\left( {{v_2}} \right) = \| {{{\bf{f}}_2}} \|_1 \vspace{1ex} \\
W\left( {{v_3}} \right) = \| {{{\bf{f}}_3}} \|_1\qquad\qquad\qquad  W\left( {{v_4}} \right) = \| {{{\bf{f}}_4}} \|_1\\
W\left( {{v_5}} \right) = {g_{{v_5}}} \cdot {\left\| {{{\bf{f}}_{{G_{{v_5}}}}}} \right\|_2} + {s_{{v_5}}} \cdot \sum\limits_{i = 1}^3 {\left| {W\left( {{v_i}} \right)} \right|} \vspace{0.7ex}\\
\quad\qquad \  = {g_{{v_5}}} \cdot {\left\| {{{\bf{f}}_{{G_{{v_5}}}}}} \right\|_2} + {s_{{v_5}}} \cdot \left( {\| {{{\bf{f}}_1}} \|_1 + \| {{{\bf{f}}_2}} \|_1 + \| {{{\bf{f}}_3}} \|_1} \right) \vspace{1ex} \\
\quad\qquad \ = {g_{{v_5}}} \cdot {\left\| {{{\bf{f}}_{{G_{{v_5}}}}}} \right\|_2} + {s_{{v_5}}} \cdot {\left\| {{{\bf{f}}_{{G_{{v_5}}}}}} \right\|_1} \vspace{1ex}\\
W\left( {{v_6}} \right) = {g_{{v_6}}} \cdot {\left\| {{{\bf{f}}_{{G_{{v_6}}}}}} \right\|_2} + {s_{{v_6}}} \cdot \left( {\left| {W\left( {{v_3}} \right)} \right| + \left| {W\left( {{v_4}} \right)} \right|} \right)\vspace{1ex}\\
 \quad\qquad \  = {g_{{v_6}}} \cdot {\left\| {{{\bf{f}}_{{G_{{v_6}}}}}} \right\|_2} + {s_{{v_6}}} \cdot \left( {\| {{{\bf{f}}_3}} \|_1 + \| {{{\bf{f}}_4}} \|_1} \right) \vspace{1ex} \\
 \quad\qquad \  = {g_{{v_6}}} \cdot {\left\| {{{\bf{f}}_{{G_{{v_6}}}}}} \right\|_2} + {s_{{v_6}}} \cdot {\left\| {{{\bf{f}}_{{G_{{v_6}}}}}} \right\|_1}\vspace{1ex}\\
W\left( {{v_{\text{root}}}} \right) =  W\left( {{v_7}} \right) = {g_{{v_7}}} \cdot {\left\| {{{\bf{f}}_{{G_{{v_7}}}}}} \right\|_2} + {s_{{v_7}}} \cdot \left( {\left| {W\left( {{v_5}} \right)} \right| + \left| {W\left( {{v_6}} \right)} \right|} \right)\vspace{1ex}\\
 \quad\qquad \  = {g_{{v_7}}} \cdot {\left\| {{{\bf{f}}_{{G_{{v_7}}}}}} \right\|_2} + {s_{{v_7}}} \cdot {g_{{v_5}}} \cdot {\left\| {{{\bf{f}}_{{G_{{v_5}}}}}} \right\|_2} + {s_{{v_7}}} \cdot {g_{{v_6}}} \cdot {\left\| {{{\bf{f}}_{{G_{{v_6}}}}}} \right\|_2} + \vspace{1ex} \\
  \quad\qquad \quad \   {s_{{v_7}}} \cdot {s_{{v_5}}} \cdot {\left\| {{{\bf{f}}_{{G_{{v_5}}}}}} \right\|_1} + {s_{{v_7}}} \cdot {s_{{v_6}}} \cdot {\left\| {{{\bf{f}}_{{G_{{v_6}}}}}} \right\|_1}\vspace{1ex} \\
\end{array}
\end{equation}
In Eq.(\ref{eqn17}), $v_1, v_2, v_3, v_4$ are the leaf nodes and the penalty terms reduce to the $l_1$ norm, which is equivalent to the $l_2$ norm in the case of a single variable. On the other hand, $v_5$ and $v_6$ are the internal nodes, so the penalty terms $W(v_5)$ and $W(v_6)$ involve both the $l_2$ norm and the $l_1$ norm. It should also be noted in Eq.(\ref{eqn17}) that $\|\mathbf{f}_{G_{v_5}}\|_1=\|\mathbf{f}_1\|_1+\|\mathbf{f}_2\|_1+\|\mathbf{f}_3\|_1=\|\mathbf{f}_1\|_2+\|\mathbf{f}_2\|_2+\|\mathbf{f}_3\|_2$ and $\|\mathbf{f}_{G_{v_6}}\|_1=\|\mathbf{f}_3\|_1+\|\mathbf{f}_4\|_1=\|\mathbf{f}_3\|_2+\|\mathbf{f}_4\|_2$. So that the weights $\omega_{v_j}$ in Eq.(\ref{eqn14}) can be represented in a compact form as follows.
\begin{equation}
  \label{eqn18}
  {\omega _{v_j}} = \left\{ \begin{array}{l}
{\prod\limits_{a \in {\rm{Ancestors}}({v_j})} {{s_a}} } \quad \quad \qquad {\text{if $v_j$ is a leaf node}} \\
{{g_{{v_j}}}\prod\limits_{a \in {\rm{Ancestors}}({v_j})} {{s_a}} }\qquad\qquad {\text{otherwise}}
\end{array} \right.
\end{equation}
The weighting scheme in Eq.(\ref{eqn18}) is a hierarchical extension of the elastic-net penalty~\cite{Kim2010}. A higher value of $s_{v_j}$ encourages the separate selection of variables in node $v_j$ and a higher value of $g_{v_j}$ favors a joint selection of the variables in $v_j$. In the extreme case of $s_{v_j}=1$ and $g_{v_j}=0$ the tree-structured penalty reduces to the LASSO penalty. On the other hand, if $s_{v_j}=0$ and $g_{v_j}=1$ the tree-structured penalty reduces to the block sparsity penalty in Eq.(\ref{eqn10}).
\section{ADMM-based solution algorithm}\label{sec:ADMM}

The fault isolation problems defined in Eqs.(\ref{eqn9}), (\ref{eqn10}), (\ref{eqn11}), (\ref{eqn13}) and (\ref{eqn14}) involve $l_1$ and $l_2$ penalty terms, making the corresponding optimization problems essentially non-convex. In solution of such constrained optimization problems, the Alternating Direction Method of Multipliers(ADMM) algorithm is a widely applied method. It combines the decomposability of the dual ascent method and the excellent convergence of the Augmented Lagrangian Method. Through decomposition and coordination, ADMM algorithm divides the objective function into several sub-optimization problems that are easy to solve, resulting in higher computational efficiency~\cite{Boyd2010}. In order to illustrate how to solve constraint optimization problems, this section presents the solution procedures for the tree-structured sparsity problem defined in Eq.(\ref{eqn14}). Solutions of other optimization problems can be obtained in a similar but simpler way.

For the sake of simplicity, the penalty term for the tree-structured sparsity in Eq.(\ref{eqn14}) is divided into two parts, each corresponding to a group of nodes in tree $T$. The first part is associated to the internal nodes and the root nodes, defined as ${V_{{\mathop{\text{ int}}} }}$, and the second part is associated to the leaf nodes ${V_{\text{leaf}}}$. So that Eq.(\ref{eqn14}) becomes
\begin{equation}
   \label{eqn19}
\tilde{\mathbf{f}}=\mathop {\min }\limits_{\bf{f}} {\left( {{\bf{x}} - {\bf{f}}} \right)^T}{\bf{M}}\left( {{\bf{x}} - {\bf{f}}} \right) + \lambda \!\!\!  \sum\limits_{v \in {V_{{\rm{int}}}}} \!\!  {{\omega _v}{{\left\| {{{\bf{f}}_{{G_v}}}} \right\|}_2}}  + \lambda \!\!\! \sum\limits_{v \in {V_{{\rm{leaf}}}}} \!\! \!{{\omega _v}{{\left\| {{{\bf{f}}_{{G_v}}}} \right\|}_2}}
\end{equation}
It should be noted that the first penalty term corresponds to the internal nodes and the root node, and the internal nodes may have overlapping variables. And the ${l_2}$ norm in the second penalty term for the leaf node ${{V_{{\text{leaf}}}}}$ is equivalent to ${l_1}$ norm as each leaf node consists of only one process variable, so the second penalty item can be expressed as $\sum\limits_{v \in {V_{{\rm{leaf}}}}} \!\!{{\omega _v}{{\left\| {{{\bf{f}}_{{G_v}}}} \right\|}_1}}$. Considering the whole sample set ${\bf{X}} \in {R^{n \times m}}$, the unconstrained optimization problem of Eq.(\ref{eqn19}) becomes
\begin{equation}
  \label{eqn20}
\tilde{\mathbf{f}}=\mathop {\min }\limits_{\bf{f}} \frac{1}{n}\sum\limits_{i = 1}^n \!\! {{{\left( { \ {\bf{x}}_i - {\bf{f}} \ } \right)}^T}{\bf{M}}\!\left( {\ {\bf{x}}_i - {\bf{f}} \ } \right)} + \lambda \! \!\! \sum\limits_{v \in {V_{{\rm{int}}}}}\!\!  {{\omega _v}{{\left\| { \ {{\bf{f}}_{{G_v}}}} \right\|}_2}}  +  \lambda  \!\!\! \sum\limits_{v \in {V_{{\rm{leaf}}}}}\!\!\!  {{\omega _v}{{\left\| { \ {{\bf{f}}_{{G_v}}}} \right\|}_1}}
\end{equation}
where ${\bf{x}}_i^T \in {^{1 \times m}}$ is the $i$th row of ${\bf{X}}$. For convenience of notation, ${{\bf{f}}_{{G_v}}}$ in the $l_2$ and $l_1$ penalty terms is reformed as ${{\bf{f}}_{{G_v}}} = {{\bf{I}}_{{G_v}}}{\bf{f}}$, where ${{\bf{I}}_{{G_v}}}\in \mathbb{R}^{m\times m}$ is a square matrix with ${{\bf{I}}_{{G_v}}}\left( {v,v} \right) = 1$ for all $v \in {G_v}$, and other elements being $0$.

Since Eq.(\ref{eqn20}) contains both the non-smooth ${l_1}$ term and the smooth ${l_2}$ term, we introduce two auxiliary variables $\left( {{{\bf{V}}_v},{\bf{Z}}} \right) \to \left( {{{{\bf{I}}_{{G_v}}}{\bf{f}}}| \ {v \in {V_{{\rm{int}}}}} ,\mathbf{f}} \right)$. Note that ${{{\bf{V}}_v}}$ corresponds to the ${l_2}$ penalty term ${{{\bf{I}}_{{G_v}}}{\bf{f}}}$, and ${\bf{Z}}$ corresponds to the ${\bf{f}}$ in the ${l_1}$ penalty term, Eq.(\ref{eqn20}) becomes
\begin{equation}
    \label{eqn21}
\begin{array}{*{20}{l}}
{\mathop {\min }\limits_{\bf{f}} \frac{1}{n}\sum\limits_{i = 1}^n {{{\left( \ {{\bf{x}}_i^T - {\bf{f}}}\ \right)}^T}} {\bf{M}}\left( { \ {\bf{x}}_i^T - {\bf{f}}} \ \right)}
{\ \ + \lambda \!\!\!\!\sum\limits_{v \in {V_{{\rm{int}}}}} \!\!\! {\omega _v}{{\left\| {\ {{\bf{V}}_v}} \ \right\|}_2} + \lambda \!\!\!\!\sum\limits_{v \in {V_{{\rm{leaf}}}}} \!\!\! {\omega _v}{{\left\| \ {{{\bf{I}}_{{G_v}}}{\bf{Z}}} \ \right\|}_1}}\vspace{1ex}\\
{\;{\rm{s}}.{\rm{t}}.\;\;{{\bf{I}}_{{G_v}}}{\bf{f}} = {{\bf{V}}_v},\;{\bf{f}} = {\bf{Z}}}
\end{array}
\end{equation}
Thus, the augmented Lagrange optimization problem is established as follows.
\begin{equation}
 \label{eqn22}
\begin{array}{l}
L\left( {{\bf{f}},{{\bf{V}}},{\bf{Z}},{\bf{U}},{\bf{R}}} \right) = \mathop {\min }\limits_{\bf{f}} \frac{1}{n}\sum\limits_{i = 1}^n {{{\left( { \ {\bf{x}}_i^T - {\bf{f}} \ } \right)}^T}{\bf{M}}\left( { \ {\bf{x}}_i^T - {\bf{f}} \ } \right)} + \lambda  \!\! \! \sum\limits_{v \in {V_{{\rm{int}}}}} \!\!\!  {{\omega _v}{{\left\| {{{ \ {\bf{V}}_v} \ }} \right\|}_2}} \vspace{1ex}\\
 \ \  \quad \quad \qquad \quad  \quad  \quad+ \ \lambda  \!\!\! \sum\limits_{v \in {V_{{\rm{leaf}}}}} \!\!\! {{\omega _v} \ {{\left\| { \ {{\bf{I}}_{{G_v}}}{\bf{Z}} \ } \right\|}_1}} + \!\!{\sum\limits_{v \in {V_{{\rm{int}}}}} \!\!{\frac{\rho }{2}\left\| { \ {{\bf{I}}_{{G_v}}}{\bf{f}} -{{\bf{V}}_v} + {{\bf{U}}_v} \ } \right\|_F^2} }  \vspace{1ex}\\
 \ \   \quad  \quad \qquad \quad  \quad  \quad+ \  \frac{\rho }{2}\left\| { \ {\bf{f}} - {\bf{Z}} + {\bf{R}} \ } \right\|_F^2  - {\frac{\rho }{2}\! \! \sum\limits_{v \in {V_{{\rm{int}}}}}\!\! {\left\| { \ {{\bf{U}}_v}} \ \right\|_F^2} } - \frac{\rho }{2}\left\| \ {\bf{R}} \  \right\|_F^2
\end{array}
\end{equation}
where ${\left\|  \cdot  \right\|_F}$ is the Frobenius norm. ${{\bf{U}}_v}$, ${{\bf{R}}}$ are scaled dual variable matrices, and $\rho  > 0$ is the ADMM regularization parameter. The core idea of ADMM iteration is to update one parameter each time (with other parameters fixed), so the optimization problem can be divided into the following series of sub-problems.
\subsection{Updating ${\bf{f}}$}
Assume the estimations of parameter matrices at the $j$th iteration have been obtained. By minimizing the term corresponding to ${{\bf{f}}}$ in  Eq.(\ref{eqn22}), the $(j+1)$th iteration ${{\bf{f}}^{(j + 1)}}$ can be obtained as follows.
\begin{equation}
  \label{eqn23}
\begin{array}{l}
{{\bf{f}}^{\left( {j + 1} \right)}} = \mathop {\arg \min }\limits_{\bf{f}} \ \frac{1}{n}\sum\limits_{i = 1}^n {{{\left( {\ {\bf{x}}_i^T - {\bf{f}}} \ \right)}^T} {\bf{M}} \left( {\ {\bf{x}}_i^T - {\bf{f}}}\ \right)} \; \vspace{1ex}\\
 \quad \ \ \ \quad + \!\!{\sum\limits_{v \in {V_{{\rm{int}}}}} \!\!{\frac{\rho }{2}\left\| { \ {{\bf{I}}_{{G_v}}}{\bf{f}} - {\bf{V}}_v^{\left( j \right)} + {\bf{U}}_v^{\left( j \right)}} \right\|_F^2} } + \frac{\rho }{2}\left\| { \ {\bf{f}} - {{\bf{Z}}^{\left( j \right)}}+{{\bf{R}}^{\left( j \right)}}}\! \right\|_F^2
\end{array}
\end{equation}
Then, differentiating Eq.(\ref{eqn23}) and setting the derivative to zero, the iterative formula of ${{\bf{f}}^{\left( {j + 1} \right)}}$ can be obtained as follows.
\begin{equation}
 \label{eqn24}
{{\bf{f}}^{\left( {j + 1} \right)}} = {\left[ {2{\bf{M}} + \rho \left( {{\bf{I}} + \!\!\!\!\sum\limits_{v \in {V_{{\rm{int}}}}} \!\!{{{\bf{I}}_{{G_v}}}} } \right)} \right]^{ - 1}}\left[ {\frac{2}{n}\sum\limits_{i = 1}^n {{\bf{M}}{{\bf{x}}_i}}  + {{\bf{D}}^{\left( j \right)}}} \right]
\end{equation}
where ${{\bf{D}}^{\left( j \right)}} = \rho \left( {\sum\limits_{v \in {V_{{\rm{int}}}}} \!\! {{{\bf{I}}_{{G_v}}}{\bf{V}}_v^{\left( j \right)}}  + {{\bf{Z}}^{\left( j \right)}} - \!\!\!\sum\limits_{v \in {V_{{\rm{int}}}}} \!\!{{{\bf{I}}_{{G_v}}}{\bf{U}}_v^{\left( j \right)}}  - {{\bf{R}}^{\left( j \right)}}} \right)$, and ${\bf{I}} \in {\mathbb{R}^{{m} \times {m}}}$ is the identity matrix.
\subsection{Updating ${\bf{V}}$}
The ${l_2}$ penalty in the objective function is controlled by ${\bf{V}}$. For each internal node ${v \in {V_{{\rm{int}}}}}$, the calculation process of ${{{\bf{V}}_v}}$ is similar to that of ${\bf{f}}$, which can be obtained ${\bf{V}}_v^{\left( {j+1} \right)}$ by minimizing all factors related to ${{{\bf{V}}_v}}$ in  Eq.(\ref{eqn22}).
\begin{equation}
 \label{eqn25}
{\bf{V}}_v^{\left( {j + 1} \right)} = \lambda \ {\omega _v}{\left\| { \ {{\bf{V}}_v} \ } \right\|_2} + \frac{\rho }{2}\left\| { \ {{\bf{I}}_{{G_v}}}{{\bf{f}}^{\left( {j + 1} \right)}} - {{\bf{V}}_v} + {\bf{U}}_v^{\left( j \right)}} \right\|_F^2
\end{equation}
The optimization problem of ${{{\bf{V}}_v}}$ can be solved by the proximal algorithm~\cite{Parikh2014}. By expressing the proximal operation target as ${\bf{B}}_v^{\left( j \right)} = {{\bf{I}}_{{G_v}}}{{\bf{f}}^{\left( {j + 1} \right)}} + {\bf{U}}_v^{\left( j \right)}$, the updating at the $(j+1)$th iteration is as follows.
\begin{equation}
  \label{eqn26}
{\bf{V}}_v^{\left( {j + 1} \right)} = \left\{ {\begin{array}{*{20}{l}}
{0\qquad \qquad \qquad \qquad \quad \;{{\left\| {{\bf{B}}_v^{\left( j \right)}} \right\|}_2} \le \frac{{\lambda {\omega _v}}}{\rho }}\vspace{1ex}\\
{\left( {1 - \frac{{\lambda {\omega _v}}}{{\rho {{\left\| {{\bf{B}}_v^{\left( j \right)}} \right\|}_2}}}} \right){\bf{B}}_v^{\left( j \right)}\qquad {\rm{otherwise}}}
\end{array}} \right.
\end{equation}
At the $(j+1)$th iteration, a numerical updating procedure will be performed for each internal node ${v \in {V_{{\rm{int}}}}}$.
\subsection{Updating ${\bf{Z}}$}
The ${l_1}$ penalty in the objective function Eq.(\ref{eqn22}) is controlled by ${\bf{Z}}$. As before, the terms containing ${\bf{Z}}$ yield a sub-problem whose objective function can be defined as follows.
\begin{equation}
 \label{eqn27}
{{\bf{Z}}^{\left( {j + 1} \right)}} = \lambda \! \! \sum\limits_{v \in {V_{{\rm{leaf}}}}} \! \!{{\omega _v}{{\left\| { \ {{\bf{I}}_{{G_v}}}{{\bf{Z}}^{\left( j \right)}}} \right\|}_1}}  + \frac{\rho }{2}\left\| { \ {{\bf{f}}^{\left( {j + 1} \right)}}\! - \!{{\bf{Z}}^{\left( j \right)}} \!+ \!{{\bf{R}}^{\left( j \right)}}} \! \right\|_F^2
\end{equation}
which can be solved by the element-wise proximal algorithm~\cite{Parikh2014}.
\begin{equation}
 \label{eqn28}
{\left[ {{{\bf{Z}}^{\left( {j + 1} \right)}}} \right]_g} = \left\{ {\begin{array}{*{20}{l}}
{0\qquad \qquad \qquad \qquad \quad  \left| {{{\bf{C}}_g}} \right| \le r}\\
{{\rm{sign}}\left( {{{\bf{C}}_g}} \right)\left( {\left| {{{\bf{C}}_g}} \right| - r} \right)\;\;\;\ {\text{otherwise}}}
\end{array}} \right.
\end{equation}
where $r = \lambda \!\!\sum\limits_{v \in {V_{{\rm{leaf}}}}} \!\!{{\omega _v}{{\left[ {{{\bf{I}}_{{G_v}}}} \right]}_g}} /\rho $, and ${\bf{C}} = {{\bf{f}}^{\left( {j + 1} \right)}} + {{\bf{R}}^{\left( j \right)}}$, with $\mathbf{C}_g$ being the $g$th element. In addition, ${\left[  \cdot  \right]_g}$ also represents the $g$th element in a vector. For a matrix, however, ${\left[  \cdot  \right]_g}$ denotes the $g$th diagonal element.
\subsection{Updating ${{\bf{U}}}$ and ${{\bf{R}}}$}
The dual variables ${{\bf{U}}}$ and ${{\bf{R}}}$ are updated by setting the derivative of the corresponding quadratic Frobenius norm to zero. The update of ${{\bf{U}}}$ needs to be performed on each internal node ${v \in {V_{{\rm{int}}}}}$, and the $(j+1)$th iteration of ${{\bf{U}}_v}$ for each internal node can be expressed as follows.
\begin{equation}
 \label{eqn29}
{\bf{U}}_v^{\left( {j + 1} \right)} = {\bf{U}}_v^{\left( j \right)} + {{\bf{I}}_{{G_v}}}{{\bf{f}}^{\left( {j + 1} \right)}} - {\bf{V}}_v^{\left( {j + 1} \right)}
\end{equation}
The scaled dual variable ${{\bf{R}}}$ is updated by
\begin{equation}
 \label{eqn30}
{\bf{R}}^{\left( {j + 1} \right)} = {\bf{R}}^{\left( j \right)} + {\bf{f}}^{\left( {j + 1} \right)} - {\bf{Z}}^{\left( {j + 1} \right)}
\end{equation}
The processing flow chart of the algorithm is summarized in Algorithm 1.

\begin{algorithm}[htbp]
\caption{ ADMM algorithm for solving Eq.(\ref{eqn14})}
\label{alg:Framwork}
\begin{algorithmic}[1]
\REQUIRE
The data matrix $\mathbf{X}$, the positive semi-definite matrix $\mathbf{M}$, weight $\omega_v$, regularization parameters $\lambda $ and $\rho $ and threshold $\varepsilon $;
\STATE\textbf{Initialize:} Randomly initialize $\mathbf{f}=\mathbf{f}^0$. Initialize auxiliary variables ${\bf{V}}$ and ${\bf{Z}}$, scaled dual variable matrices ${\bf{U}}$ and ${\bf{R}}$ as zero vector/matrix, and set $j = 0$;\\
\STATE\textbf{Do until convergence}\\
\quad\quad 1) Updating ${{\bf{f}}}$£»\vspace{-1.5mm}

\quad\quad 2) Updating ${{\bf{V}}}$£»\vspace{-1.5mm}

\quad\quad 3) Updating ${{\bf{Z}}}$£»\vspace{-1.5mm}

\quad\quad 4) Updating ${{\bf{U}}}$ and ${{\bf{R}}}$\\

\STATE\textbf{if}\ $\left\| {\frac{{{{\bf{f}}}^{\left( {j + 1} \right)} - {{\bf{f}}}^{\left( j \right)}}}{{{{\bf{f}}}^{\left( {j + 1} \right)}}}} \right\| > \varepsilon$, set $j = j + 1$ and return to step 2;\vspace{1ex}
\STATE\textbf{end}
\STATE\textbf{Return ${{\bf{f}}}$, ${{\bf{V}}}$ and ${{\bf{Z}}}$}
\end{algorithmic}
\end{algorithm}

\subsection{Determination of weight and regularization parameters}
In order to obtain the solution of each structured sparse objective function, the parameters $\lambda $, $\rho $ and $\omega_v$ should be determined in advance. Based on the discussion in~\cite{Yan2015}, there exists a finite sequence of transition points ${\lambda _0} > {\lambda _1} >  \cdots {\lambda _K} = 0$ such that the induced sparse result changes at each transition point. when $\lambda  > {\lambda _0}$, contributions of all variables shrink to zero, i.e. ${\bf{f}} = 0$ and the active set ${\rm A} = \emptyset $. When $\lambda  < {\lambda _K}$ , all variables are selected and ${\rm A} = \left\{ {1,2, \cdots ,m} \right\}$. Within each interval $\left( {{\lambda _h},{\lambda _{h + 1}}} \right)$, the active set ${\rm A}$ does not change. Therefore, the choice of $\lambda $ can be based on the transition points, because they induce the best solution under the same sparsity. In order to select an appropriate $\lambda$ from the active set, the reconstructed monitoring statistic is compared with the corresponding control limits. The value that produces the optimum reconstruction is selected. In order to speed up the parameter determination step, the candidate $\lambda$ values are selected from the interval of $\left[\sqrt{\frac{\log m}{n}}, \pi \sqrt{\frac{\log m}{2n}}\right]$, following the suggestion of Ref.~\cite{Boyd2010}. For the parameter $\rho$, the range of $\rho  = 1.2 \sim 1.5$ has been recommended in Boyd et al.~\cite{Boyd2010}. And for the $\omega_v$ in the tree-structured sparsity, it is related to two quantities $g_v$ and $s_v$, which are associated with the height $h_v$ of each node $v$ in the tree $T$, given as ${s_v} = {h_v}$ and ${g_v} = 1 - {h_v}$. The height $h_v$ can be defined as the number of edges in the longest path from node $v$ down to a certain leaf node~\cite{Huang2019}, indicating how tightly its members are correlated. In addition, it is assumed that the height $h_v$ of each node is normalized so that the height of the root node is 1. For the tree structure in Figure \ref{Figure2}, we can get ${h_{{v_1}}}{\rm{ = }}{h_{{v_2}}}{\rm{ = }}{h_{{v_3}}}{\rm{ = }}{h_{{v_4}}}=0$, ${h_{{v_5}}}{\rm{ = }}{h_{{v_6}}}{\rm{ = }}0.5$, ${h_{{v_7}}}{\rm{ = }}1$.

\section{Simulation studies}
This section tests the performance of the structured sparsity modeling methods on fault isolation of a simulation example involving 15 process variables. The variables are assumed to fall into 4 natural blocks, as is shown in Table~\ref{table1}.
\begin{table}[htbp]
\begin{center}
\small
  \caption{Natural blocks of process variables in the simulation studies.}
    \label{table1}
  \begin{tabular}{ccccc}
    \\[-13mm]
    \hline
    \hline\\[-9mm]
   \quad \ {\bf  Block No.}& \quad \quad 1 & \quad \quad 2 &\quad \quad 3 &\quad \quad 4 \quad \ \\
    \hline
    \vspace{1mm}\\[-9mm]
   \quad \ {\bf  Indices of process variables}& \quad \quad 1, 2, 6, 7, 10  & \quad \quad 3, 11, 15 & \quad \quad 4, 9, 13  &  \quad\quad 5, 8, 12, 14  \quad \ \\
    \hline
    \hline
  \end{tabular}
  \end{center}
\end{table}
The variables in all the blocks are assumed to be generated from one or two source variables as follows.
\begin{equation} \label{eqn31}
\begin{array}{l}
{\rm{Block1}}\left\{ \begin{array}{l}
{{\bf{x}}_1} \sim N(0,1)\\
{{\bf{x}}_2} \sim N(0,1)\\
{{\bf{x}}_6}{\rm{ = }}0.6{{\bf{x}}_1} + 0.4{{\bf{x}}_2} + 0.03\epsilon \\
{{\bf{x}}_7}{\rm{ = }}0.4{{\bf{x}}_1} + 0.3{{\bf{x}}_2} + 0.3{{\bf{x}}_6} + 0.02\epsilon\\
{{\bf{x}}_{10}}{\rm{ = }}0.2{{\bf{x}}_1} + 0.5{{\bf{x}}_2} + 0.1{{\bf{x}}_6} + 0.2{{\bf{x}}_7} + 0.01\epsilon
\end{array} \right. \! \!\!\!\!\!\! \!\!\!\!\!\! {\rm{Block2}}\left\{ \begin{array}{l}
{{\bf{x}}_3} \sim N(0,1)\\
{{\bf{x}}_{11}}{\rm{ = }}0.8{{\bf{x}}_3} + 0.03\epsilon\\
{{\bf{x}}_{15}}{\rm{ = }}0.3{{\bf{x}}_3} + 0.7{{\bf{x}}_{11}} + 0.01\epsilon
\end{array} \right.\\
{\rm{Block3}}\left\{ \begin{array}{l}
{{\bf{x}}_4} \sim N(0,1)\\
{{\bf{x}}_9}{\rm{ = }}0.7{{\bf{x}}_4} + 0.02\epsilon\\
{{\bf{x}}_{13}}{\rm{ = }}0.6{{\bf{x}}_4} + 0.4{{\bf{x}}_9} + 0.02\epsilon
\end{array} \right.
{\rm{Block4}}\left\{ \begin{array}{l}
{{\bf{x}}_5} \sim N(0,1)\\
{{\bf{x}}_8}{\rm{ = }}0.8{{\bf{x}}_5} + 0.02\epsilon\\
{{\bf{x}}_{12}}{\rm{ = }}0.5{{\bf{x}}_5} + 0.5{{\bf{x}}_8} + 0.03\epsilon\\
{{\bf{x}}_{14}}{\rm{ = }}0.2{{\bf{x}}_5} + 0.4{{\bf{x}}_8} + 0.4{{\bf{x}}_{12}} + 0.01\epsilon
\end{array} \right.\quad
\end{array}
\end{equation}
Here $\epsilon \sim N(0,1)$ is the Gaussian distributed noise with zero mean and unit variance. It can be seen from Eq.(\ref{eqn31}) that the source variables for the first block are $x_1$ and $x_2$, and those for the remaining blocks are $x_3$, $x_4$ and $x_5$ respectively.

For the purpose of model training, a total of 700 samples are generated as the training sample set. An additional 300 samples are generated as the test set, which involves a sensor bias and a multiplicative fault introduced after the 101st sample. Using the training data, a PCA model is constructed. It is found that retaining 5 principal components(PCs) retains more than 85\% variance so that the number of PCs is set as 5. Based on the PCA model, the $T^2$ and SPE statistics can be constructed to monitor the process, with the control limits obtained under the significance level of 0.01.

\begin{figure}[htbp]
\begin{center}
\includegraphics[width=10cm]{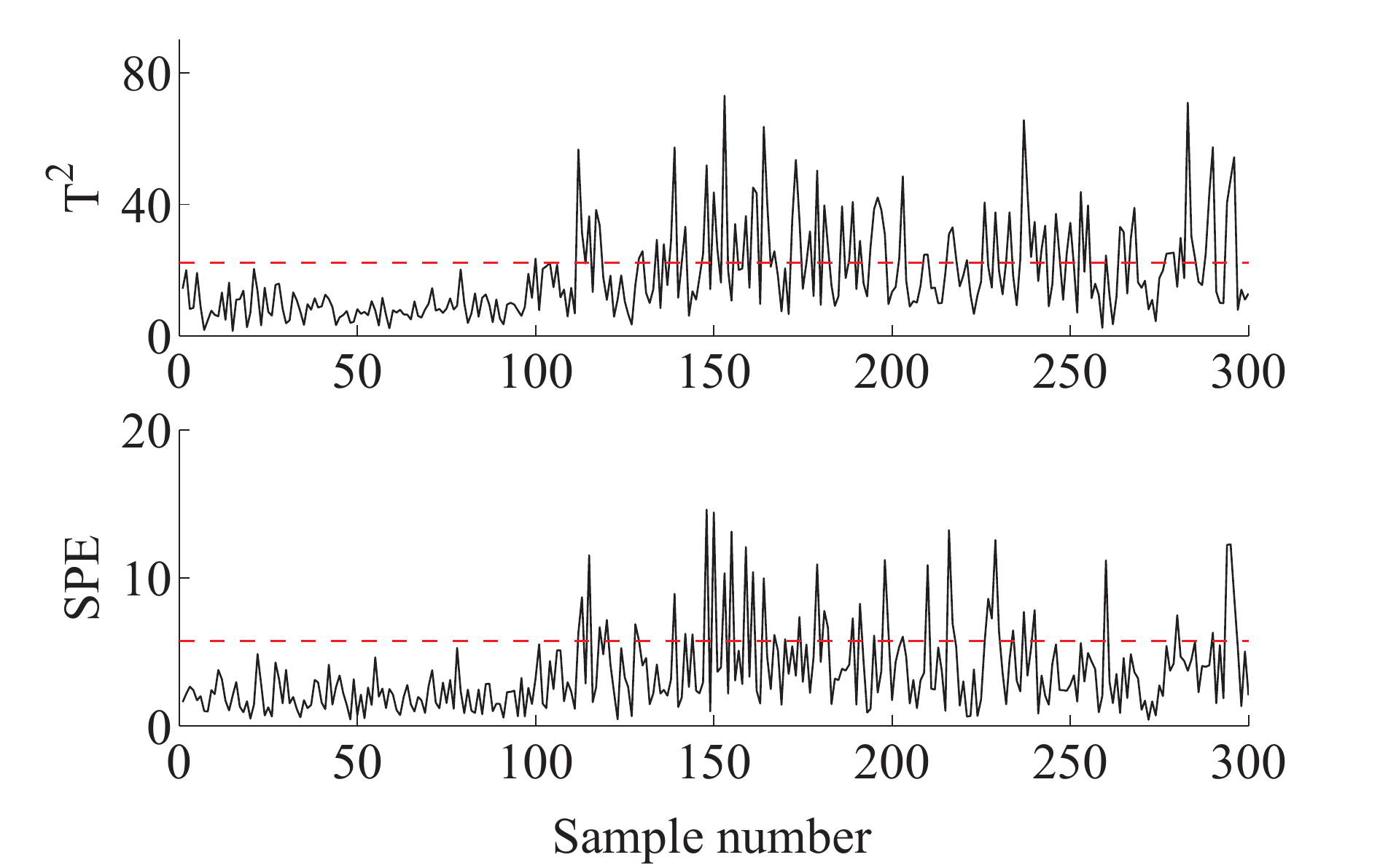}
\setlength{\abovecaptionskip}{-3pt}
\caption{PCA based monitoring statistics for the faulty samples involving the sensor bias.}  \label{Figure3}
\end{center}
\end{figure}

For the sensor bias, the fault is introduced on variable $x_7$ as $\hat{x}_7=x_7-1.5$. The monitoring results for the test set is shown in Figure~\ref{Figure3}. It can be seen that significant number of violations can be observed for both the $T^2$ and SPE statistics after the 101st sample, indicating a fault has occurred. This is in accordance with our previous setting. After the fault is successfully detected, the next important task is to isolate the faulty variables and diagnose the fault type.

\begin{figure}[htbp]
\begin{center}
\includegraphics[width=15cm]{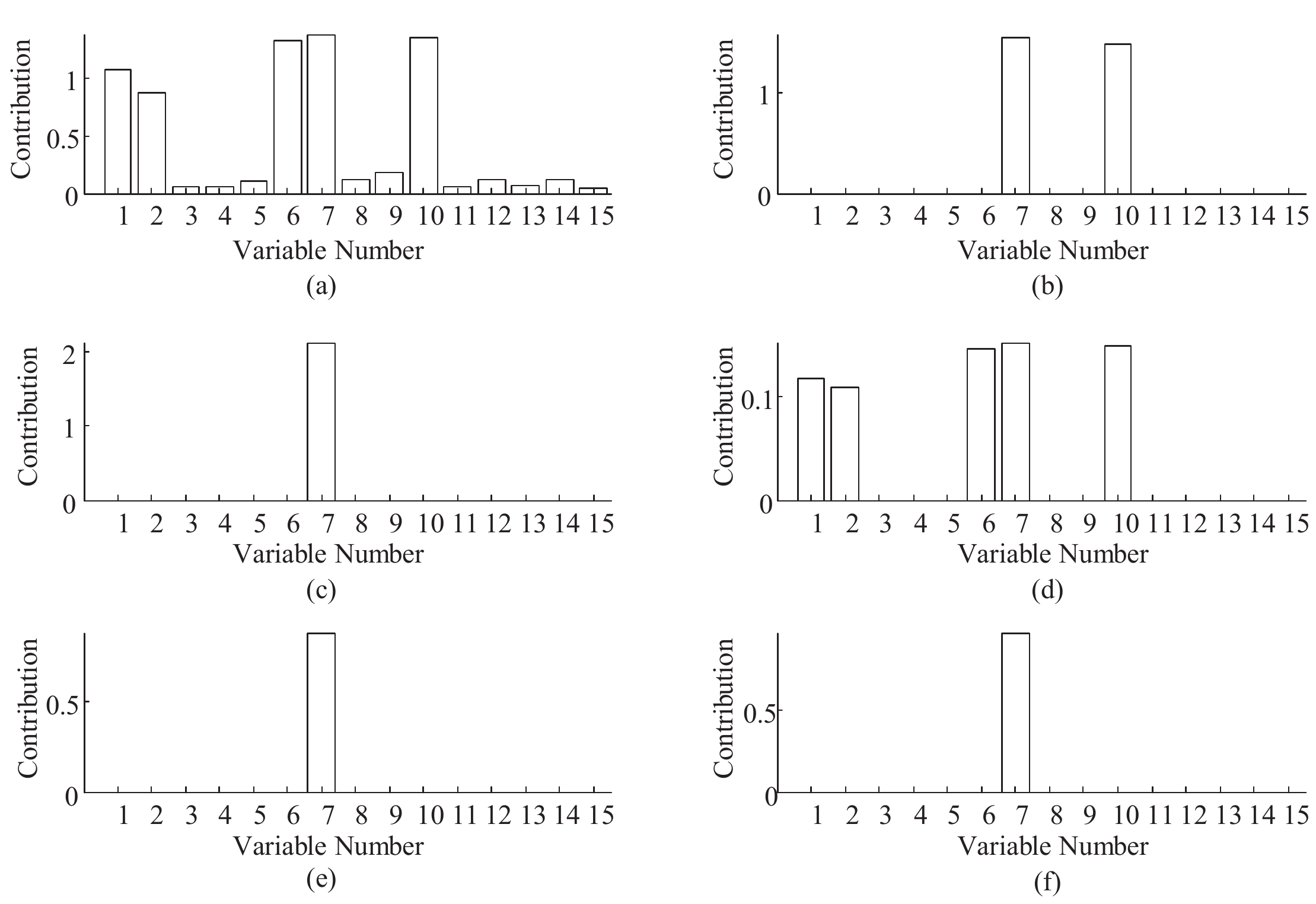}
\setlength{\abovecaptionskip}{-3pt}
\caption{Isolation results of different methods for the sensor bias: (a) Conventional reconstruction based contribution plot ; (b) only the $l_1$ penalty considered; (c) partially known sparse support; (d) Group Lasso; (e) Sparse Group Lasso; (f) Tree-structured sparsity.}  \label{Figure4}
\end{center}
\end{figure}

Before performing fault isolation, it is essential to obtain the block structure of variables. Among the four structured sparsity terms, the blocks of variables for block sparsity, clustered sparsity and tree-structured sparsity are obtained from analyzing normal operation data or the natural relationship between process variables. For the structured sparsity term of partially known sparse support, it requires knowledge of the faulty behavior of the process as it needs to predetermine a subset of faulty/normal variables. In this example, the block structures of Group Lasso, sparse Group Lasso and tree-structured sparsity are obtained from process knowledge, whilst partially known sparse support is obtained from knowledge about faulty behavior. By incorporating the process structure using the methods proposed in Section 3, the fault isolation results for the sensor bias are shown in Figure~\ref{Figure4}, which is obtained based on the 200 identified faulty samples. Figure~\ref{Figure4} consists of six plots, corresponding to the fault isolation results obtained from Eqs.(\ref{eqn6}), (\ref{eqn7}), (\ref{eqn9}), (\ref{eqn10}), (\ref{eqn11}) and (\ref{eqn14}) respectively. Note that clustered sparsity is not considered here as the block structure is clear in this example. The parameter $\lambda$ is set as 0.6, 0.6, 1.0, 0.9 and 0.8 in Eqs.(\ref{eqn7}), (\ref{eqn9}), (\ref{eqn10}), (\ref{eqn11}) and (\ref{eqn14}) respectively by applying the regularization parameter determination method proposed in Section 4.5. The regularization parameter $\rho$ in the ADMM algorithm is set as 1.2, and the pre-determined threshold $\varepsilon $ is set as ${10^{ - 6}}$.
In Figure~\ref{Figure4}(a), in the absence of any additional information, the fault isolation results obtained using conventional reconstruction based contribution successfully identified the fault variable $x_7$, however, the fault contributions in Figure~\ref{Figure4}(a) also show high contributions in $x_1, x_2, x_6, x_{10}$. This is due to the ``smearing effect''~\cite{Shang2019}. In addition, the obtained contribution vector is not a sparse one, making it difficult to distinguish normal variables from faulty ones. In Figure~\ref{Figure4}(b), the $l_1$ penalty term is introduced and a sparse contribution vector can be obtained, however, the contribution of variables $x_{10}$ cannot be neglected. This can be explained, as it has been proved that LASSO-like fault isolation method cannot fully overcome the ``smearing effect''. In Figure ~\ref{Figure4}(c), it is assumed that $x_{10}$ in Block 1 is known to be normal so that partial support information is known. Comparing Figures~\ref{Figure4}(b) and ~\ref{Figure4}(c) it can be seen that after introducing the partial support information, the fault isolation accuracy has been increased significantly. In Figure~\ref{Figure4}(d), the block information is considered and group LASSO is used to identify the faulty group.
\begin{figure}[htbp]
\begin{center}
\includegraphics[width=10cm]{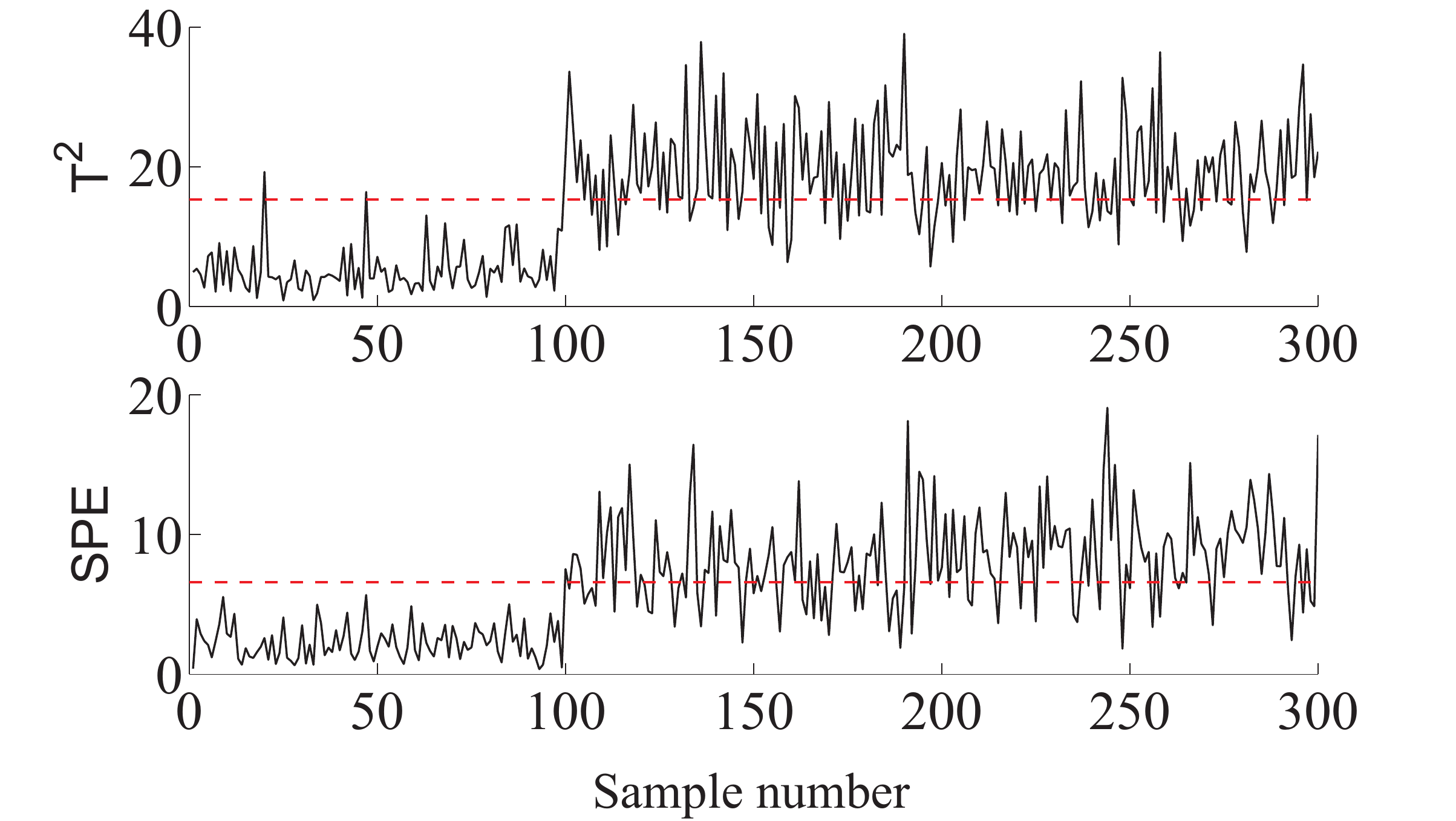}
\setlength{\abovecaptionskip}{-3pt}
\caption{PCA based monitoring results for the faulty samples involving the multiplicative fault.}  \label{Figure5}
\end{center}
\end{figure}
As is shown in Figure~\ref{Figure4}(d),  faulty blocks are successfully identified as the contributions from variables in Blocks 1 are not zero. However, as is pointed out in Section 3.2, solving the optimization problem in Eq.(\ref{eqn10}) will result in block-wise sparsity, making the task of further pinpoint the fault source difficult. In Figure~\ref{Figure4}(e), the sparse group Lasso is used to identify fault variables. Comparing Figure~\ref{Figure4}(d) and Figure~\ref{Figure4}(e), it is found that introducing sparsity inside the blocks can better locate the fault source. Finally, using the correlation structure in the four blocks, a tree-structure can be obtained, as shown in Figure~\ref{Figure7}. Based on the tree-structure, the fault isolation problem in Eq.(\ref{eqn14}) is solved, and the results are shown in Figure~\ref{Figure4}(f). From Figure~\ref{Figure4}(f) it can be seen that by utilizing the tree-structure, the variable $x_7$ is correctly identified.
Comparing the six plots in Figure~\ref{Figure4} it can be seen that by introducing structured sparsity information, the fault isolation accuracy can be significantly improved and the developed structured sparsity methods are effective.

\begin{figure}[htbp]
\begin{center}
\includegraphics[width=15cm]{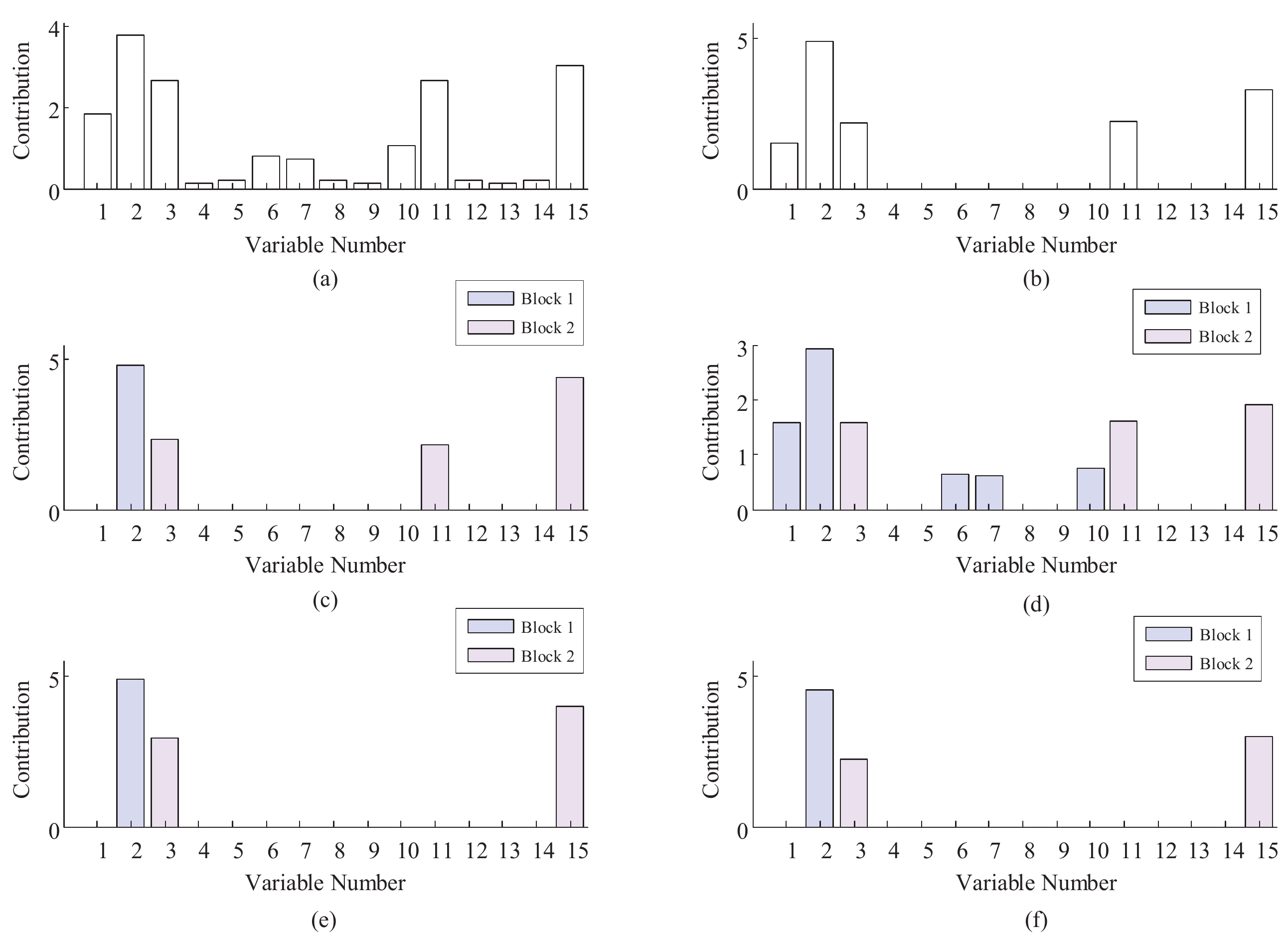}
\setlength{\abovecaptionskip}{-3pt}
\caption{Isolation results of various methods for the multiplicative fault: (a) Conventional reconstruction based contribution plot ; (b) only the $l_1$ penalty considered; (c) partially known sparse support; (d) Group Lasso; (e) Sparse Group Lasso; (f) Tree-structured sparsity.}  \label{Figure6}
\end{center}
\end{figure}

\begin{figure}[htbp]
\begin{center}
\includegraphics[width=12cm]{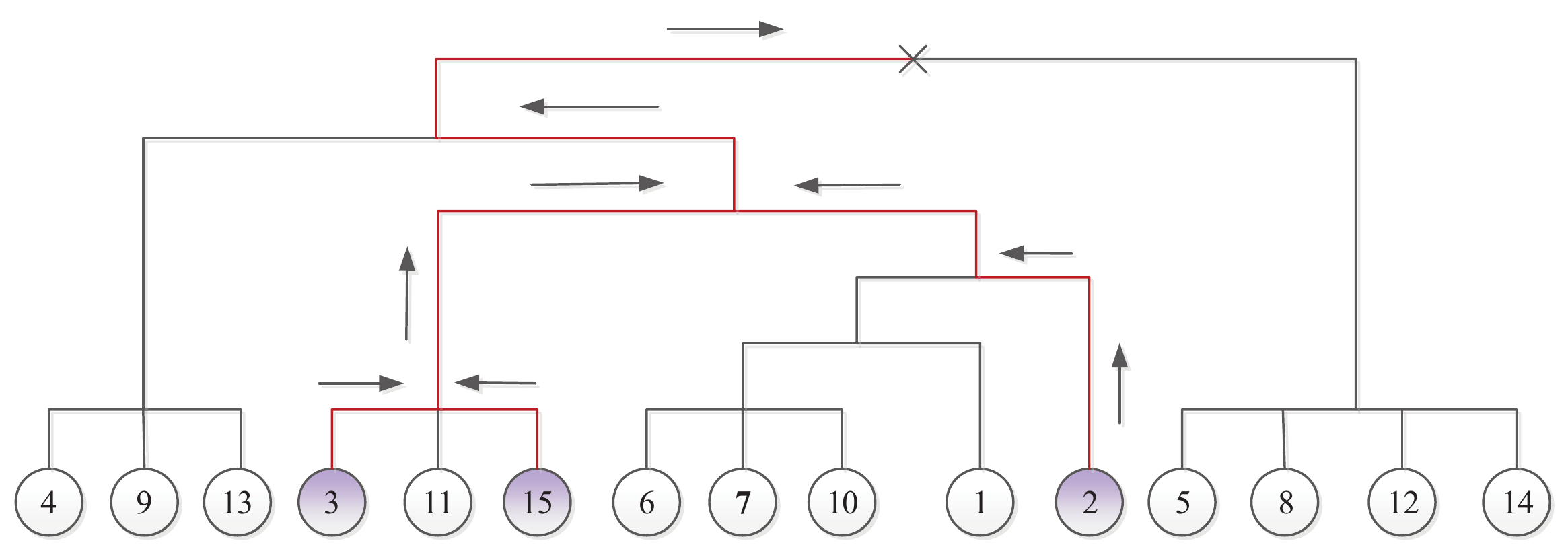}
\caption{The tree structure of the multiplicative fault and the identified fault propagation path.}  \label{Figure7}
\end{center}
\end{figure}

For the simulation of multiplicative fault, the fault is generated on variables $x_2, x_3, x_{15}$ by multiplying a scalar as $\hat{x}_2=0.5x_2, \hat{x}_3=0.8x_3, \hat{x}_{15}=0.6x_{15}$. The PCA model is constructed in the same way as the case of sensor fault, and the process monitoring statistics for the test set are shown in Figure~\ref{Figure5}. As designed, it is found that significant number of violations are observed after the 101st sample, indicating a fault is observed. Based on the 200 identified faulty samples, the fault isolation results of various methods are shown in Figure~\ref{Figure6}. The parameter $\lambda$ is set as 0.6, 0.7, 1.4, 1.2 and 1.0 in Eqs.(\ref{eqn7}), (\ref{eqn9}), (\ref{eqn10}), (\ref{eqn11}) and (\ref{eqn14}) respectively. The regularization parameter $\rho$ and the pre-determined threshold $\varepsilon $ remain unchanged.

In Figure~\ref{Figure6}(a), $x_2, x_3, x_{15}$ are successfully identified as fault variables, but due to the `smearing effect'', the variables $x_1, x_{10}, x_{11}$ also have non-ignorable effects. In Figure~\ref{Figure6}(b), the $l_1$ penalty term is introduced, but the contribution of variables $x_1, x_{11}$ cannot be ignored. In Figure~\ref{Figure6}(c), assuming that $x_{1}$ is normal, the accuracy of the result is improved compared to Figure~\ref{Figure6}(b), but $x_{11}$ is still mistaken as faulty. In Figure~\ref{Figure6}(d), the group LASSO identified all variables in Block 1 and Block 2 as faulty, making it difficult to locate the fault source. By introducing sparsity in the group, Figure~\ref{Figure6}(e) successfully identified $x_2, x_3, x_{15}$ as faulty. Finally, using the tree structure in Figure~\ref{Figure7}, the fault diagnosis result based on the tree-structured sparsity is shown in Figure~\ref{Figure6}(f). From Figure~\ref{Figure6}(f), it can be seen that the three faulty variables $x_2, x_3, x_{15}$ are also correctly identified. In addition, the tree structure in Figure~\ref{Figure7} clearly indicates the fault propagation path, which is helpful for the operators in taking corrective operations.

\section{Application in coal-fired power plant}
This section applies the proposed structured sparsity method to isolation of a fault in the coal pulverizer of a coal-fired power plant. The coal-fired power plant generates steam with high temperature and high pressure by burning coal. The generated steam then drives the steam turbine to rotate, which further drives the generator to generate electricity. The coal-fired power plant relies on a series of critical components to convert the energy stored in the coal ore into electricity, and one of which is the coal pulverizer. During the operation of a coal pulverizer, the coal is fed into the pulverizer, crushed by the grinder, then dried and classified by the air using the separator to ensure only fine powdered coal particles are sent to the boiler for combustion. As a critical device in the coal-fired power plant, it is important to accurately detect and isolate any fault in the coal pulverizer to ensure there is enough time for the operators to make corrective operations. The coal pulverizer has a clear operating mechanism and process structure, which can be useful for fault isolation.
\begin{figure}[htbp]
\begin{center}
\includegraphics[width=14.5cm]{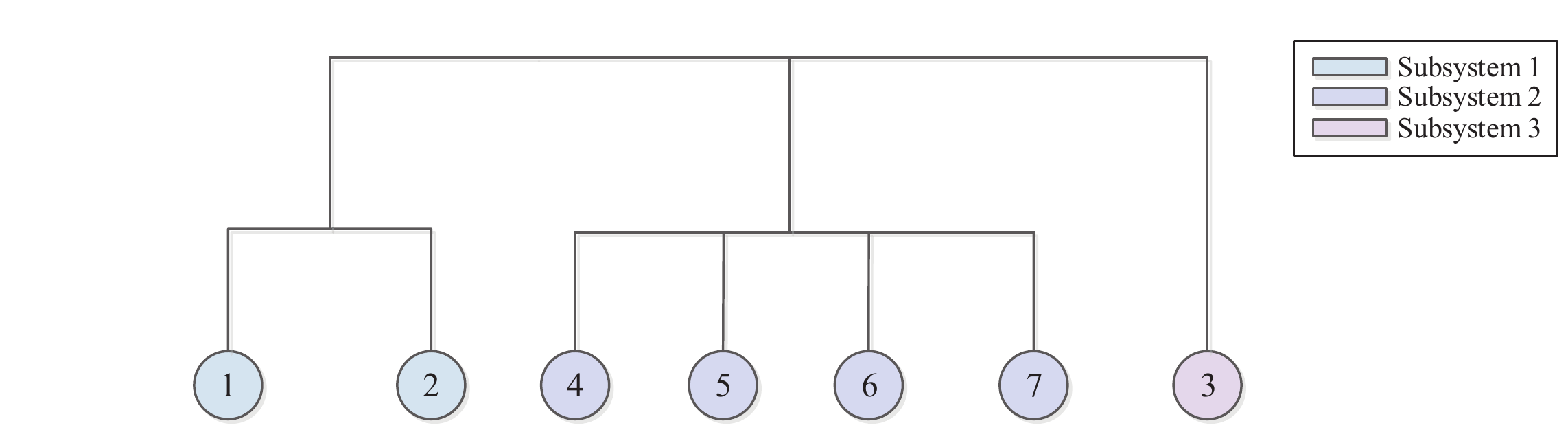}
\setlength{\abovecaptionskip}{-3pt}
\caption{Coal pulverizer tree structure.}  \label{Figure8}
\end{center}
\end{figure}

\begin{table}[htbp]
\begin{center}
\footnotesize
  \caption{Process variables and description for the coal pulverizer.}
    \label{table2}
  \begin{tabular}{cccc}
    \\[-4mm]
    \hline
    \hline\\[-9mm]
     {\bf Variable No.}&   {\bf  Description}  &  {\bf Variable No.}& {\bf  Description}  \\
  \hline
    \vspace{1mm}\\[-9mm]
    1-4 &  Inlet air flow(1-4) &  26 &  Hot primary air baffle valve position \\
    5&  Inlet air pressure&  27 &  Separator current \\
     6-8  &    Inlet air temperature(1-3) &  28 &  Separator frequency\\
     9-13 &    Coal feed(1-5)&   29&  Separator speed\\
    14-15 &   Outlet powdered coal pressure(1-2)&   30-35 &  Motor stator temperature(1-6)\\
   16-22 & Outlet powdered coal temperature(1-7)&   36-37 & Motor bearing temperature(1-2)\\
   23 &  Cold primary air baffle control instruction&   38-40 &  Roller bearing oil temperature(1-3)\\
   24&  Cold primary baffle valve position &41-42 &  Thrust bearing tile temperature(1-2)\\
   25&  Hot primary air baffle control instruction&43-46 &  Gearbox oil sump temperature(1-4)\\
        \hline
         \hline
  \end{tabular}
  \end{center}
\end{table}

For the purpose of model verification, a dataset collected from a coal-fired power plant in southeast China is considered. The dataset consists of a training set with 3000 normal samples and a test set with 1000 samples. The test set involves an abnormal operation caused by decreased outlet powered coal temperature in the last 600 samples. The decreased outlet powered coal temperature caused the control system to turn down the cold primary baffle valve to reduce the cold air feed, which further caused the grinder to reduce the workload. A total of 46 variables($x_1$ through $x_{46}$) related to the coal pulverizer is considered and described in Table~\ref{table2}.

Similar to Section 5, the block structures for Group Lasso, sparse Group Lasso and tree-structured sparsity are obtained by natural relationship between process variables. And the block structure for partially known sparse support is determined from knowledge of the faulty behavior. For the coal pulverizer, the 46 variables can be naturally divided into 7 blocks corresponding to process inputs, process outputs, primary air baffle, separator, electric motor, grinder and gearbox, as is shown in Table~\ref{table3}. The relationship between the 7 blocks can be summarized using the tree structure shown in Figure~\ref{Figure8}.

\begin{table}[htbp]
\begin{center}
\footnotesize
  \caption{ Natural blocks of process variables in the coal pulverizer.}
    \label{table3}
  \begin{tabular}{ccc}
      \\[-4mm]
    \hline
    \hline\\[-8mm]
    \quad \ \quad \ \quad \  {\bf Block No.} \quad \ & \quad \  \quad \  {\bf  Description} \quad \  \quad \ & \quad \  {\bf Indices of process variables} \quad \  \quad \ \quad \ \\
  \hline
    \vspace{1mm}\\[-9mm]
  \quad \  \quad \  \quad \ 1  \quad \ &  \quad \ \quad \ Input  \quad \ \quad \ & \quad \  1, 2, 3, 4, 5, 6, 7, 8, 9, 10, 11, 12, 13  \quad \ \quad \ \quad \ \\
  \quad \ \quad \   \quad \ 2  \quad \ & \quad \ \quad \  Output \quad \ \quad \ & \quad \  14, 15, 16, 17, 18, 19, 20, 21, 22 \quad \ \quad \ \quad \ \\
    \quad \ \quad \  \quad \ 3 \quad \ & \quad \ \quad \ Primary air baffle  \quad \ \quad \ & \quad \   23, 24, 25, 26 \quad \ \quad \ \quad \ \\
   \quad \ \quad \  \quad \ 4  \quad \ & \quad \ \quad \ Separator \quad \ \quad \ & \quad \  27, 28, 29 \quad \ \quad \ \quad \ \\
   \quad \ \quad \  \quad \ 5  \quad \ & \quad \ \quad \   Electric motor \quad \ \quad \ & \quad \  30, 31, 32, 33, 34, 35, 36, 37 \quad \ \quad \ \quad \ \\
   \quad \  \quad \  \quad \ 6 \quad \ & \quad \ \quad \ Grinder \quad \ \quad \ &  \quad \  38, 39, 40, 41, 42 \quad \ \quad \ \quad \ \\
   \quad \ \quad \  \quad \ 7 \quad \ &  \quad \ \quad \   Gearbox \quad \ \quad \ &  \quad \  43, 44, 45, 46 \quad \ \quad \ \quad \ \\
        \hline
         \hline
  \end{tabular}
  \end{center}
\end{table}

\begin{figure}[htbp]
\begin{center}
\includegraphics[width=10cm]{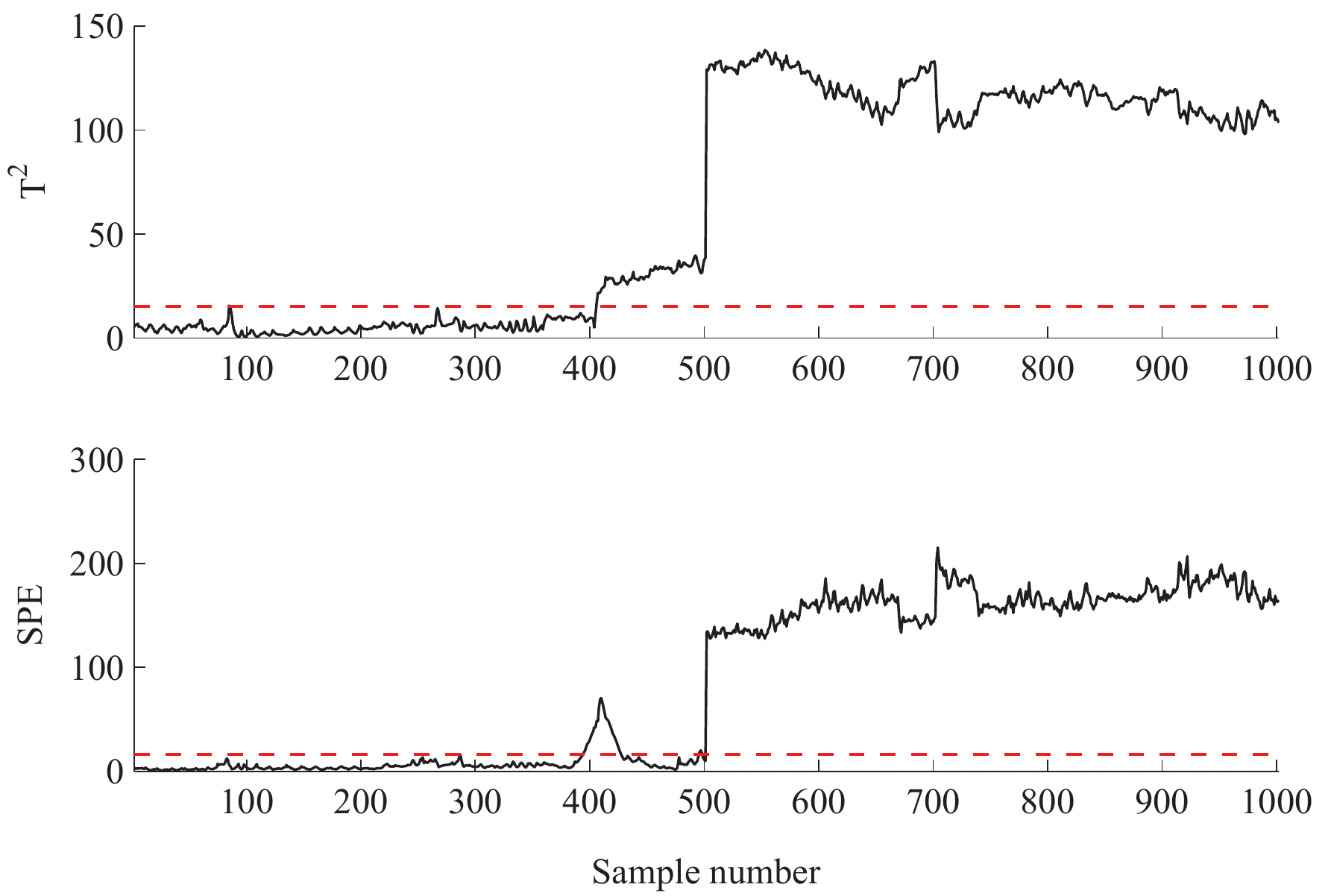}
\setlength{\abovecaptionskip}{-3pt}
\caption{PCA-based monitoring statistics for the 1000 test samples.}  \label{Figure9}
\end{center}
\end{figure}

In order to detect the fault, the 3000 training samples are used to build a PCA-based monitoring model and it is found that retaining 5 PCs can explain more than 85\% variance, so that the number of retained PCs is set as 5. By setting a significance level of 0.01, the monitoring results are obtained and shown in Figure~\ref{Figure9}.

It can be seen that significant number of violations can be observed in both the $T^2$ and SPE statistics after the 400th sample point, indicating a fault is detected. After the fault is detected, fault isolation is then performed to identify faulty variables using the structured sparsity method.

\begin{figure}[t]
\begin{center}
\includegraphics[width=15.5cm]{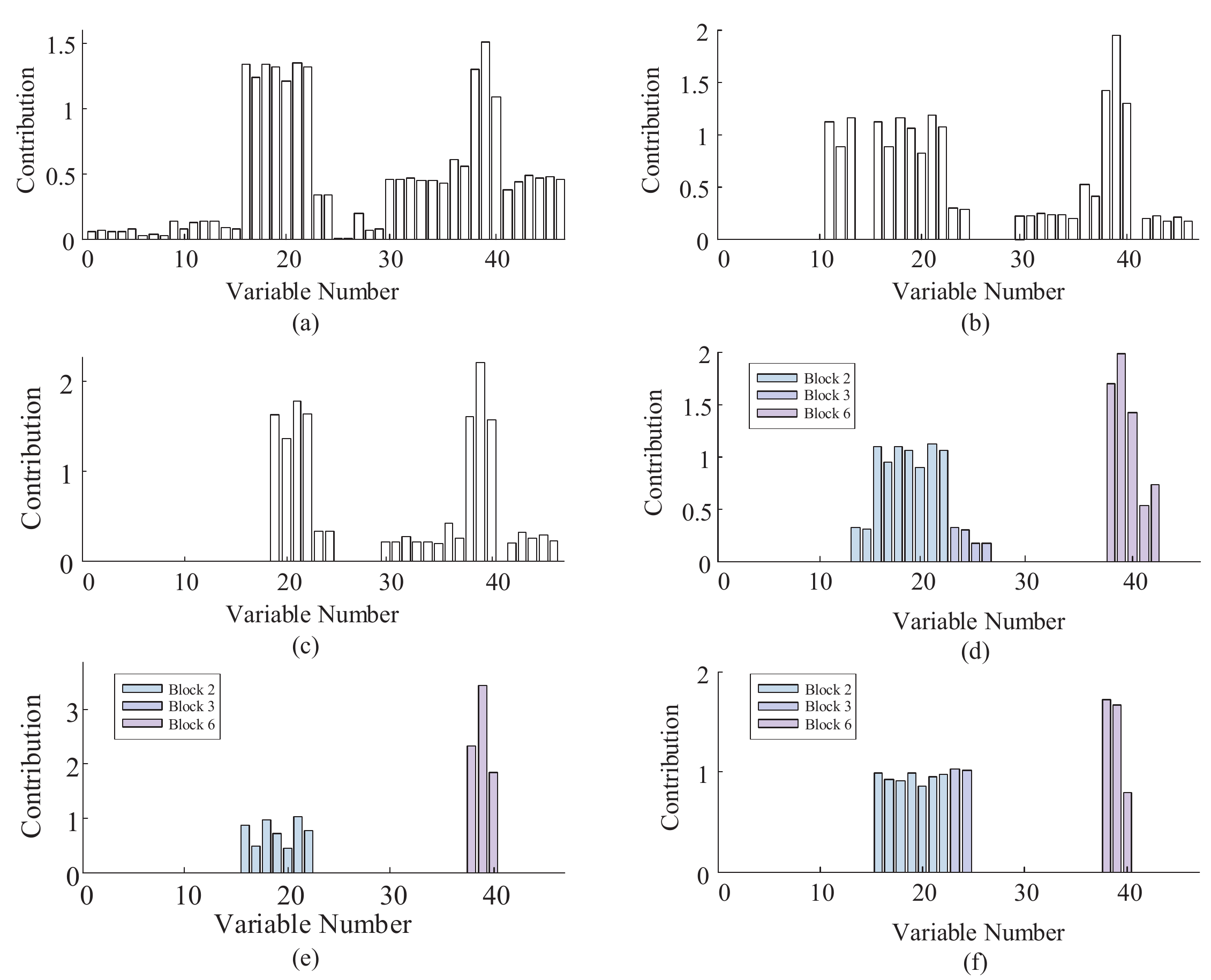}
\caption{Comparison of fault isolation results using different approaches: (a) Conventional reconstruction based contribution plot ; (b) only the $l_1$ penalty considered; (c) partially known sparse support; (d) Group Lasso; (e) Sparse Group Lasso; (f) Tree-structured sparsity.}
\label{FIGURE10}
\end{center}
\end{figure}

Six kinds of methods are applied and compared, including the conventional reconstruction based contribution plot, the $l_1$ sparsity, partially known sparse support, group LASSO, sparse group LASSO and the tree-structured sparsity, and the fault reconstruction results are shown in Figure~\ref{FIGURE10}. The fault isolation results in Figure~\ref{FIGURE10} are obtained based on the 600 identified faulty samples. The parameter $\lambda$ for the $l_1$ penalty and partially known sparse support based method are set as 0.5, while in Eq.(\ref{eqn10}), Eq.(\ref{eqn11}) and Eq.(\ref {eqn14}), the parameter $\lambda$ is set to 0.65, 0.8, 1.2 respectively, the ADMM regularization parameter $\rho$ and the pre-determined parameter $\varepsilon $ are the same as in the simulation. It can be seen in Figure~\ref{FIGURE10}(a) that conventional reconstruction based contribution plot identified plenty of variables to be faulty, making it difficult to locate the fault source. This is expected, as the reconstruction based contribution plot suffers from the ''smearing effect''. Also, as is shown in Figure~\ref{FIGURE10}(b), the $l_1$ penalty based method identified variables  ${x_{11}} \sim {x_{13}}$, ${x_{16}} \sim {x_{22}}$, and ${x_{38}}\sim {x_{40}}$ to be faulty, and helps to narrow down the fault cause. In Figure~\ref{FIGURE10}(c), it is assumed that ${x_{11}} \sim {x_{13}}$ are known to be normal so that partial support information is known. Comparing with Figure~\ref{FIGURE10}(b), it can be seen that after introducing the partial support information, the accuracy of fault isolation is improved.

\begin{figure}[htbp]
\begin{center}
\includegraphics[width=10cm]{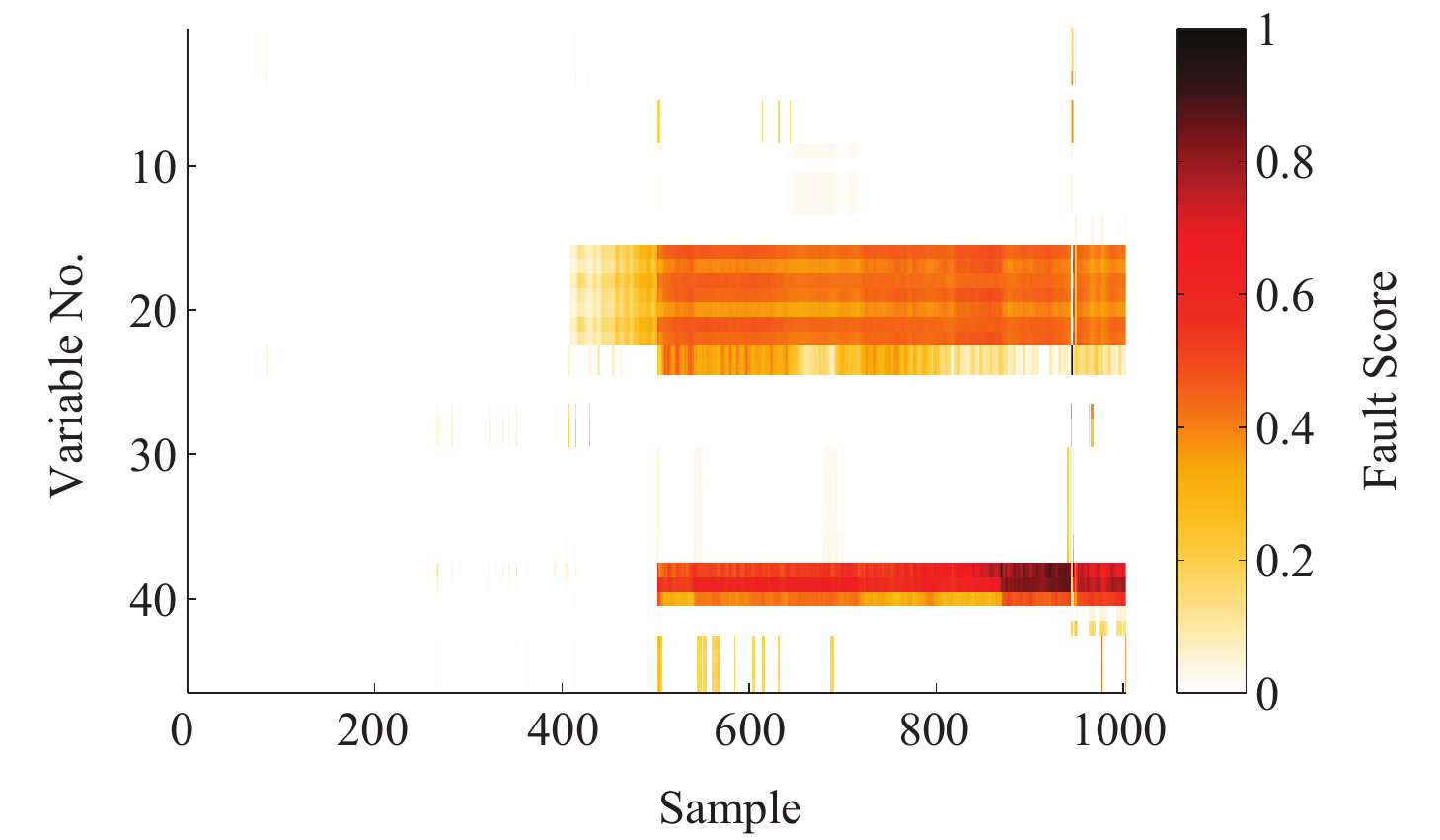}
\setlength{\abovecaptionskip}{-3pt}
\caption{Sample-by-sample fault scores using tree-structured sparsity.}  \label{Figure11}
\end{center}
\end{figure}

By introducing the group sparsity, Figure~\ref{FIGURE10}(d) correctly identified 3 blocks to be faulty, including the output block, grinder block and the primary air baffle block. However, it still lacks more specific information on which variables in each block are responsible for the fault. In Figure~\ref{FIGURE10}(e), compared to the Figure~\ref{FIGURE10}(d), it successfully located the faulty variables in Block 2 and Block 6, however, variables ${x_{23}},  {x_{24}}$ in Block 3 are identified as normal. In contrast, it can be seen from Figure~\ref{FIGURE10}(f) that the tree structured sparsity also identified the 3 blocks as faulty. In addition, it identified $x_{16}\sim x_{22}$ in the output block, $x_{23}, x_{24}$ in the primary air baffle block and $x_{38} \sim x_{40}$ in the grinder block to be faulty. That is to say, the fault affects the outlet powdered coal temperatures, the cold primary air baffle control instruction and valve position and the roller bearing oil temperatures. This can be explained, as the outlet powdered coal temperatures drop, the control system reduced the cold primary air flow as well as the workload of the roller. In order to have a more clearer picture of the fault, Figure~\ref{Figure11} gives the sample by sample fault isolation results obtained using the tree-structured sparsity method. As is shown in Figure~\ref{Figure11}, variables $x_{16}\sim x_{22}$ are first identified to be faulty, and after around 100 samples, $x_{23}, x_{24}$ as well as $x_{38}\sim x_{40}$ are also identified to be faulty. This is in accordance with the operator's later findings that the fault was caused by an abnormal operation caused by decreased outlet powered coal temperature.

\begin{figure}[htbp]
\begin{center}
\includegraphics[width=15.5cm]{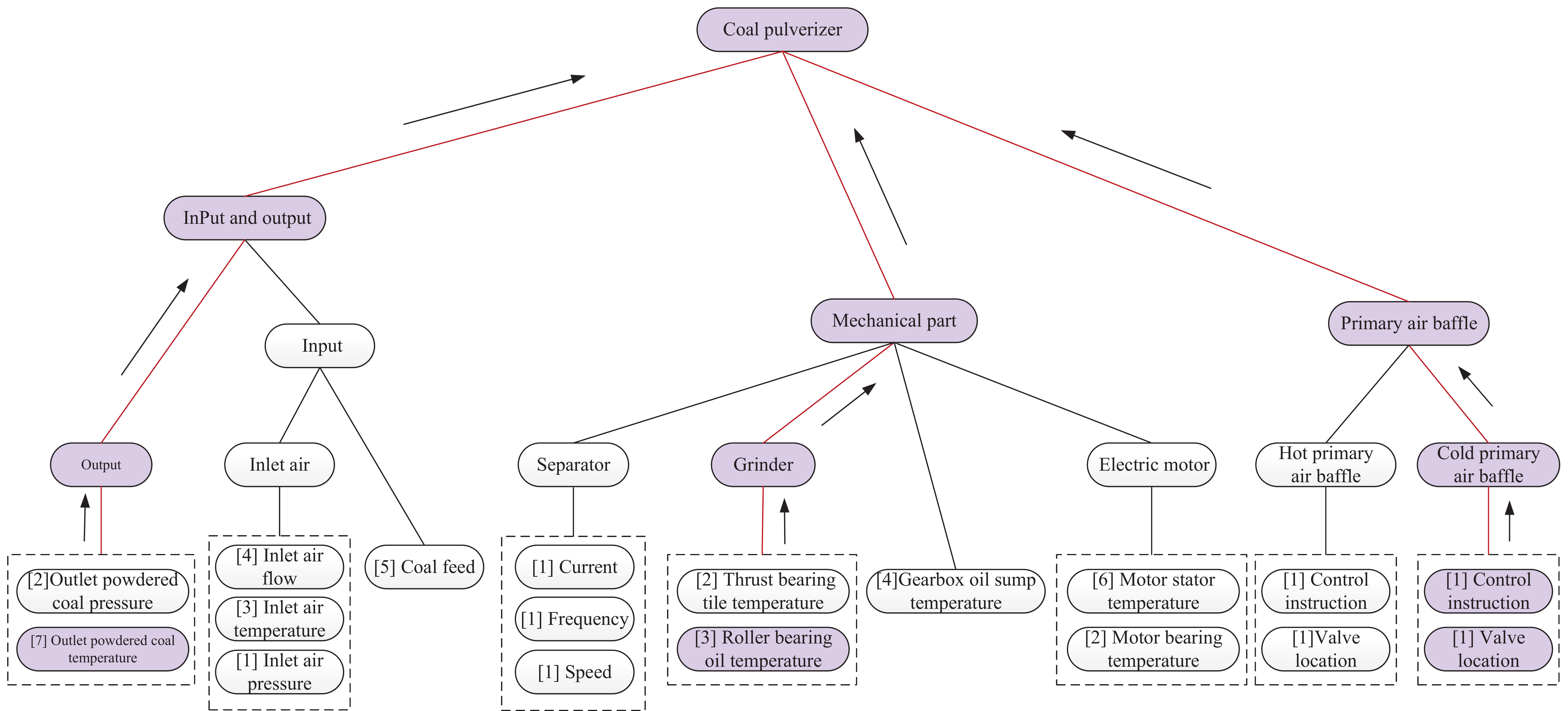}
\caption{Identified faulty structure of the coal pulverizer.}
\label{Figure12}
\end{center}
\end{figure}

Moreover, Figure~\ref{Figure12} illustrates the identified faulty variables in the tree structure, which clearly identified the affected blocks and variables. Such clear tree-structure is helpful for operators to identify the fault cause so that corrective operations can be taken in time. This clearly shows the advantages brought by the proposed structured sparsity methods.

\section{Conclusion}
This paper proposes a modeling method based on structured sparsity for fault isolation in industrial processes. The fault isolation problem is based on the conventional reconstruction based contribution analysis and four structure sparsity regularization terms are considered to incorporate process structure. It is found that these structured sparsity terms allow selection of faulty variables over different process structures. In order to solve the reconstruction problem, an algorithm based on ADMM is proposed. Through application studies to a simulation example and a fault in the coal-fired power plant, it is verified that by including the structured sparsity terms, better fault isolation accuracy can be obtained. In future research, the methods considered in this paper can be extended to deal with dynamic processes using an augmented variable matrix. In addition, to improve the evaluation of the fault diagnosis results, some kinds of objective and quantitative measures should be considered.

\section*{Acknowledgment}
The authors would like to acknowledge financial support from the Natural Science Foundation of China (Nos. 61673358, 61973145) and the NSFC-Zhejiang Joint Fund for the Integration of Industrialization and Informatization, China (U1709215).

\bibliographystyle{plain}

\bibliography{Structured_sparsity_modeling_framework}

\end{document}